\documentclass[aps,prb,twocolumn,amsmath,amssymb,superscriptaddress]{revtex4-1}
\usepackage[colorlinks=true,allcolors=blue]{hyperref}
\usepackage{amsfonts,times,bbold,dsfont}
\usepackage{graphicx}
\let\vec\boldsymbol
\providecommand{\vect}[1]{{\boldsymbol{#1}}}

\begin{document}

\title{Spin-transfer torque-driven motion, deformation, and instabilities of magnetic skyrmions at high currents}

\author{J. Masell \href{https://orcid.org/0000-0002-9951-4452}{\includegraphics[height=0.75em]{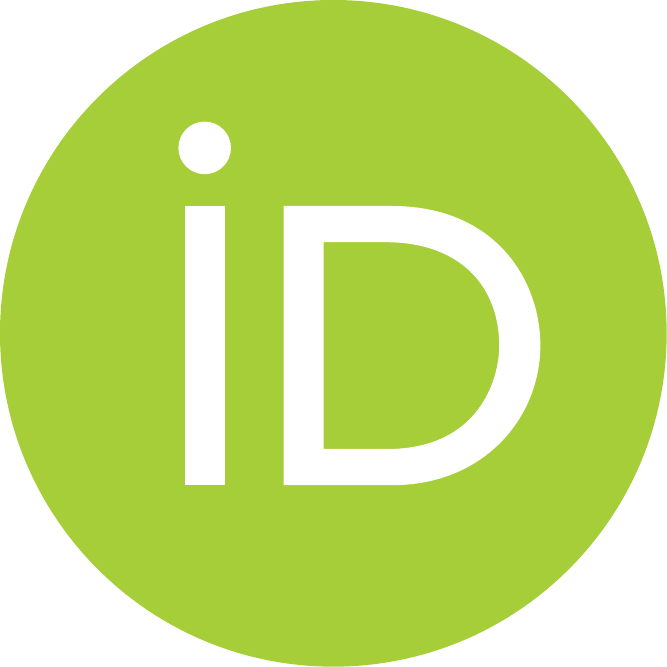}}}
\affiliation{Institute for Theoretical Physics, University of Cologne, 50937 Cologne, Germany}
\affiliation{RIKEN Center for Emergent Matter Science, Wako, Saitama 351-0198, Japan}
\author{D.R. Rodrigues \href{https://orcid.org/0000-0002-6301-4974}{\includegraphics[height=0.75em]{orcid.pdf}}}
\affiliation{Institute of Physics, Johannes Gutenberg-Universit{\"a}t, 55128 Mainz, Germany}
\affiliation{Graduate School Materials Science in Mainz, Staudingerweg 9, 55128 Mainz, Germany}
\author{B.F. McKeever \href{https://orcid.org/0000-0002-3361-6480}{\includegraphics[height=0.75em]{orcid.pdf}}}
\affiliation{Institute of Physics, Johannes Gutenberg-Universit{\"a}t, 55128 Mainz, Germany}
\affiliation{Graduate School Materials Science in Mainz, Staudingerweg 9, 55128 Mainz, Germany}
\author{K. Everschor-Sitte \href{https://orcid.org/0000-0001-8767-6633}{\includegraphics[height=0.75em]{orcid.pdf}}}
\affiliation{Institute of Physics, Johannes Gutenberg-Universit{\"a}t, 55128 Mainz, Germany}
\date{\today}

\begin{abstract}
In chiral magnets, localized topological magnetic whirls, magnetic skyrmions, can be moved by spin polarized electric currents.
Upon increasing the current strength, with prospects for high-speed skyrmion motion for spintronics applications in mind, isolated skyrmions deform away from their typical circular shape.
We analyze the influence of spin-transfer torques on the shape of a single skyrmion, including its stability upon adiabatically increasing the strength of the applied electric current.
For rather compact skyrmions at uniaxial anisotropies well above the critical anisotropy for domain wall formation, we find for high current densities that the skyrmion assumes a non-circular shape with a tail, reminiscent of a shooting star.
For larger and hence softer skyrmions close to the critical anisotropy, in turn, we observe a critical current density above which skyrmions become unstable.
We show that above a second critical current density the shooting star solution can be recovered also for these skyrmions.
\end{abstract}

\maketitle

\section{Introduction}
\label{sec:introduction}

Since the experimental discovery in chiral magnets~\cite{Muhlbauer2009a,Yu2010} and layered magnetic systems,~\cite{Heinze2011,Boulle2016} magnetic skyrmions~\cite{Bogdanov1989,Bogdanov1994,Leonov2016a,Nagaosa2013} have inspired a variety of possible applications in information technology devices.~\cite{Nagaosa2013,Jiang2017,Everschor-Sitte2018}
Their non-trivial topology is often associated with high stability against thermal fluctuations and material defects.
Similarly to magnetic domain walls in racetrack-type memories,~\cite{Parkin2008} skyrmions can be moved by spin-polarized electric currents~\cite{Jonietz2010,Schulz2012,Iwasaki2013,Iwasaki2013a, Woo2016,Jiang2017,Litzius2017} (spin currents) which suggests their use in shift-register-like devices.~\cite{Fert2013,Muller2017}
However, a type of Walker breakdown~\cite{Schryer1974,Malozemoff1979} is known to limit the speed of domain walls~\cite{Thiaville2005} by exciting the internal degrees of freedom.
An important topic for the use of skyrmions in future technological devices is the dynamic behavior of the whirls subject to high current-densities and their intrinsic upper-speed limit.\cite{Caretta2018, Gobel2019, Litzius2020}

Whether for the spin-current-driven switching of magnetic domains in MRAM elements,~\cite{Bhatti2017} or for the current-driven motion of magnetic states like domain walls~\cite{Thiaville2005} or skyrmions,~\cite{Komineas2015a,Litzius2017} two distinct mechanisms are usually considered:
(i) spin-transfer torques (STTs) for smooth magnetic textures, which arise as the spin-polarization of the conduction electrons follows almost adiabatically the direction of the magnetization and, hence, exerts a torque on the magnetization wherever it is non-collinear,~\cite{Bazaliy1998,Zhang2004,Stiles2006} and
(ii) spin-orbit torques (SOTs), which appear as effective local field-like or damping-like torques due to spin-accumulation at interfaces.~\cite{Manchon2019}
Both mechanisms are known to move ferromagnetic skyrmions not precisely in the direction of the applied current, but with an additional perpendicular velocity component.
This feature defines the skyrmion Hall angle which arises intrinsically due to a topologically nontrivial skyrmion winding number.
Since most applications consider the motion of skyrmions in constricted geometries,~\cite{Iwasaki2013a} a first limitation therefore arises from the skyrmion Hall effect\cite{Litzius2020} which can push the skyrmion out of the system.\cite{Everschor-Sitte2018, Gobel2019}
Proposals have therefore been made to avoid the perpendicular component of motion entirely, e.g., by (i) fine-tuning the polarization of the electrons for the SOT by using ferromagnetic layers with certain crystal symmetries,~\cite{Huang2017a,Kim2018, Gobel2019} (ii) adjusting the direction of the current for the STT, (iii) creating antiferromagnetically coupled layers in which the opposite winding numbers and, hence, opposite Hall effects cancel,~\cite{Zhang2016} or (iv) by solely considering antiferromagnetic systems~\cite{Bogdanov2002} where the skyrmion Hall effect is naturally absent.~\cite{Zhang2016b,Barker2016,Velkov2016}
A second challenge arises due to the presence of defects such as impurities or edges in the system which cause local deformations of the magnetization and can even act as nucleation hotspots for skyrmions if applied currents are strong enough.~\cite{Everschor-Sitte2017,Stier2017}
In addition to these practical difficulties, it has been experimentally shown that the in-plane field-like SOTs not only tilt the easy axis of the magnetization but also cause deformations of skyrmions.~\cite{Litzius2017,Juge2019, Litzius2020}
It was therefore argued that deformations due to SOTs might explain the current-dependence of the skyrmion's Hall effect~\cite{Litzius2017,Juge2019, Litzius2020} and, eventually, are responsible for their destruction.

For isolated skyrmions driven by STTs, distortions as for SOT-driven skyrmions are usually neglected.
This assumption is supported by a plethora of micromagnetic simulations which confirm that skyrmions move as approximately circular, rigid objects with speed proportional to the magnitude of applied current and at an angle to the current direction.~\cite{Iwasaki2013}
This behavior is faithfully described by the widely used Thiele method,~\cite{Thiele1973} and its extensions for example in the case of pinning.~\cite{Everschor2012,Schutte2014a,Mueller2015}
Moreover, it was shown by Lin, Ref.~\citenum{Lin2017}, that the STT-induced deformations are indeed small, using a linear response approach for skyrmions stabilized in external magnetic fields.
Larger deformations and even instabilities, in turn, are only reported for the combination of extremely large STTs and small skyrmions of the order of the atomic lattice.~\cite{Lin2013}
However, a detailed analysis of the STT-induced distortion of skyrmions does, to the best of our knowledge, not exist in the literature.

In this paper, we systematically study the distortion of isolated skyrmions by STTs with both high-precision numerical simulations and analytical approximations of the non-linear sigma model, Sec.~\ref{sec:model}.
In Sec.~\ref{sec:domainwalls}, we first review STT-driven motion and pair-annihilation of 1D domain walls for comparative purposes.
In Sec.~\ref{sec:bfield}, in agreement with Ref.~\citenum{Lin2017}, we find that skyrmions with external magnetic fields are very rigid and not much affected by STTs.
Skyrmions which are stabilized by a strong uniaxial anisotropy are, however, more attractive for device applications since they do not require invasive external magnetic fields for stability. 
We intensively study these systems in Sec.~\ref{sec:skyrmion} and find that deformations cannot be neglected and, eventually, trigger an elliptical instability of the skyrmions even for small STTs.
Moreover, we demonstrate the possible existance of current-stabilized skyrmion solutions in a regime above the elliptical instability.

\section{The model}
\label{sec:model}

To study the deformation of skyrmions due to STTs, we focus our analysis on an idealized two-dimensional system in which the magnetization is described by a non-linear sigma model.
We write the energy of a magnetic texture $\vect{m}=\vect{M}/M_s$ with respect to the out-of-plane polarized state with $\vect{m}_{\mathrm{fm}}=\hat{\vect{z}}$ as
\begin{equation}
\begin{split}
    E[\vect{m}] = z_{0}\! \!\int\!\! \mathrm{d}x \,\mathrm{d}y  \,\,
     \big[ A (\nabla \vect{m})^2 - D \,\vect{m} \cdot \left(\left(\hat{\vect{z}}\!\times\!\nabla\right)\!\times\!\vect{m}\right) \\
     - \mu_0 M_s H (m_z-1) - K (m_z^2-1) \big],
\end{split}
\label{eq:model:energy}
\end{equation}
where $M_s$ is the saturation magnetization, $A$ the magnetic stiffness, $D$ the interfacial Dzyaloshinskii-Moriya (DM) interaction, $H$ an external magnetic field and $z_{0}$ the thickness of the magnetic thin film.
Here, $K=K_u-\mu_0 M_s^2/2>0$ is an effective uniaxial anisotropy which corrects the lattice anisotropy $K_u$ by a local approximation of the magnetostatic interactions.~\cite{Bogdanov1994}
The interfacial form of the DM interaction in Eq.~\eqref{eq:model:energy} stabilizes non-collinear N\'eel-type spirals and skyrmions, i.e., where the plane of spin rotation is spanned by propagation vector $\vect{q}$ and the out-of-plane direction $\hat{\vect{z}}$ for spirals, or within planes spanned by the radial vector and $\hat{\vect{z}}$ for skyrmions.

In this paper, we will mainly focus on the dynamics of skyrmions stabilized without an external field, $H=0$.
We briefly discuss the stability diagram of a field stabilized skyrmion in Sec.~\ref{sec:bfield},
where we show that even right at the phase transition from the polarized ferromagnet to the skyrmion lattice the deformation of the skyrmion is minimal, in agreement with previous results.~\cite{Schutte2014a,Lin2017}
The reason for the small deformation is that the eigenmodes of the skyrmion have a large gap~\cite{Lin2014,Schutte2014} even close to the transition from the polarized background to a skyrmion lattice state, making these skyrmions very stiff.
Without an external field, however, internal eigenmodes of the skyrmion soften at the phase transition from the polarized state to the helicoidal phase.~\cite{Lin2014,Kim2014d}
In Sec.~\ref{sec:skyrmion}, we will therefore focus on the regime near the critical anisotropy~\cite{Bogdanov1994} $K\gtrsim K_c=\pi^2 D^2 / 16 A$ where we also observe the strongest deformations.

The dynamics of the magnetization far below the Curie temperature is well described by the Landau-Lifshitz-Gilbert (LLG) equation.~\cite{Gilbert2004}
The interaction with the electric current is included by STTs for smooth magnetic textures:~\cite{Bazaliy1998,Zhang2004}
\begin{equation}
\begin{split}
\dot{\vect{m}} =
    &- \gamma\, \vect{m} \times \tilde{\vect{B}}_{\mathrm{eff}} - \left(\vect{v}_e \cdot \nabla\right) \vect{m} \\
    &+ \alpha\, \vect{m} \times \dot{\vect{m}} + \beta\, \vect{m}\times( \vect{v}_e \cdot \nabla ) \vect{m} \,\,.
 \end{split}
\label{eq:model:LLG}
\end{equation}
where $\dot{\vect{m}} = d\vect{m}/{dt}$, $\gamma$ is the (positive) gyromagnetic ratio and $\tilde{\vect{B}}_{\mathrm{eff}}=-\delta E[\vect{m}] / (M_{s}\delta\vect{m})$ is the effective magnetic field due to interactions in the magnetization.
The dimensionless constants $\alpha$ and $\beta$ are respectively the Gilbert damping and the non-adiabatic damping parameter.
The drift velocity here is~\cite{Zhang2004} $\vect{v}_e = - [P\mu_\mathrm{B}/e M_{s}(1+\beta^{2})]\vect{j}_{e}$ with $\vect{j}_e$ the electric current density, $P$ the polarization, $\mu_\mathrm{B}$ the Bohr magneton, and $e>0$ the electron charge.

\subsection{Dimensionless unit system}
\label{sec:model:parameters}

\begin{table}[t]
\begin{center}
$
\begin{array}{ccc}  \hline
\text{Name} & \multicolumn{1}{p{2cm}}{\centering \text{Dimensionless} \\ \text{quantity}} & \multicolumn{1}{p{1cm}}{\centering \text{Dimension-full} \\ \text{replacement}}  \\    [\medskipamount]  \hline \noalign{\medskip}
\text{length} & x & (|D|/2A)x   \\
\text{time} & t & (\gamma D^{2}/2AM_{s})t  \\
\text{velocity} & v & (M_{s}/\gamma |D|)\,|\vect{v}_{e}| \\
\text{reduced anisotropy} & \kappa & 2AK/D^{2}  \\ 
\text{magnetic field} & h &  (2A M_{s}/D^{2}) \, H \\ [\medskipamount] \hline
\end{array}
$
\end{center}
\caption{
Guide to change from dimensionless quantities in final results back into usual notations with simple replacements, for estimating any measurable quantities.
The damping parameters $\alpha$ and $\beta$ do not change.
}
\label{tab:units}
\end{table}

In the remainder of this paper we will use a dimensionless unit system for convenience.
Within the continuum approximation we can express the energy in units of $2A z_0$ and length in units of $2A/D$.\footnote{For definiteness we fix $D>0$ from this point onward.}
Consequently the dimensionless effective magnetic field $\vect{B}_{\mathrm{eff}}=(2AM_{s}/D^{2})\tilde{\vect{B}}_{\mathrm{eff}}$,
\begin{align}
\vect{B}_{\mathrm{eff}}= \nabla^{2}\vect{m}  +2(\hat{\vect{z}}\times\nabla)\times\vect{m} + 2\kappa m_{z}\hat{\vect{z}},
	\label{eq:dimensionless-B-eff}
\end{align}
depends only on a single coupling constant, a dimensionless anisotropy $\kappa=2AK/D^2$.
The $\nabla$ operator in Eq.~\eqref{eq:dimensionless-B-eff} and from this point forwards is dimensionless.
In these dimensionless units the critical anisotropy separating the mean field ground states is $\kappa_{c}=\pi^{2}/8$.
In the LLG, Eq.~\eqref{eq:model:LLG}, we can absorb the prefactor of the precession term by defining a dimensionless time $\gamma D^2 (2 A M_s)^{-1} t$ and dimensionless drift velocity $\vect{v}=M_s(\gamma D)^{-1} \vect{v}_e$.
This brings the LLG to a form where only $\kappa,\alpha,\beta$ and $\vect{v}$ appear as parameters when it is written out fully in terms of the magnetization and its derivatives:
\begin{equation}
\begin{split}
\dot{\vect{m}}
&= -\vect{m}\times \vect{B}_{\mathrm{eff}} - (\vect{v}\cdot\nabla)\vect{m}
	 \\
&\quad
	+ \alpha\vect{m}\times\dot{\vect{m}}  + \beta\vect{m}\times(\vect{v}\cdot\nabla)\vect{m}.
 \end{split}
 	\label{eq:dimensionless-LLG}
\end{equation}
As a guide we provide a conversion dictionary in Table~\ref{tab:units} to change all results back into usual notations.

\subsection{Reduced sets of parameters for steady-state motion}
\label{subsec:ParameterSet}

Upon assuming to reach a steady-state motion with constant dimensionless drift velocity $\vect{v}_{d}$,  magnetic textures may be described by the traveling wave ansatz $\vect{m}(\vect{r}-\vect{v}_{d}t)$.~\cite{Thiele1973}
By transforming the LLG Eq.~\eqref{eq:dimensionless-LLG} with the steady-state assumption, we obtain (now in dimensionless units)
\begin{align}
\vect{0}
&= -\vect{m}\times \vect{B}^{\mathrm{ss}}_{\mathrm{eff}},
\label{eq:new-static-equation}
\end{align}
where the steady-state effective field $\vect{B}^{\mathrm{ss}}_{\mathrm{eff}}$  is
\begin{align}\label{eq:EffectiveFieldGeneral}
\vect{B}^{\mathrm{ss}}_{\mathrm{eff}}
& =\vect{B}_{\mathrm{eff}}  + \vect{B}_{\mathrm{DL}}+ \vect{B}_{\mathrm{FL}},
\end{align}
with
\begin{subequations}
\begin{align}
\label{eq:BDL}
\vect{B}_{\mathrm{DL}} &=\left[(\alpha\vect{v}_{d}-\beta\vect{v})\cdot\nabla\right] \vect{m}\\
\label{eq:BFL}
\vect{B}_{\mathrm{FL}} &= \vect{m}\times \left[(\vect{v}_{d}-\vect{v})\cdot\nabla\right] \vect{m}.
\end{align}
\end{subequations}
Eq.~\eqref{eq:new-static-equation} is a second order differential equation in the spatial variables only and characterizes the magnetization profile in steady-state motion due to STTs.
The current-induced terms are responsible for the deformations of these magnetic textures, compared to the magnetization profile in the absence of applied current.
Note that the equation contains precisely the effective velocity vectors appearing in the Thiele equation~\cite{Thiele1973} for translationally invariant models,
\begin{align}
\vect{\mathcal{G}}\times (\vect{v}_{d}-\vect{v}) + \mathcal{D}(\alpha\vect{v}_{d}-\beta\vect{v}) = \vect{0}.
\label{eq:Thiele-equation}
\end{align}
The first term is the Magnus force, where $\vect{\mathcal{G}}=4\pi \mathcal{Q}\hat{\vect{z}}$ is the gyrovector responsible for the Hall angle of magnetic states with a non-trivial 2D Pontryagin index
\begin{align}
\mathcal{Q} = \frac{1}{4\pi}\int dx\, dy \,\, \vect{m}\cdot(\partial_{x}\vect{m}\times \partial_{y}\vect{m}) \in \mathbb{Z},
\end{align}
while the second term is dissipative and controlled by the symmetric dissipative matrix defined by
\begin{equation}
\mathcal{D}_{ij} = \int dx \,dy\,\, \partial_{i}\vect{m}\cdot\partial_{j}\vect{m},
\end{equation}
where $i, j \in\{x,y\}$ and $\mathcal{D}_{iz}=\delta_{iz}$.
We shall later utilize the dissipative matrix $\mathcal{D}$ for characterizing current-induced deformations of topologically nontrivial magnetic textures.
To conveniently compare the differences between domain walls and skyrmions, we analyze how the reduced anisotropy coupling $\kappa$, drive parameter $\vect{u}$, and Gilbert damping $\alpha$,~\cite{Komineas2015a}
\begin{equation}
\kappa=\frac{2AK}{D^2},\,\, \quad \vect{u}=\frac{\alpha\!-\!\beta}{\alpha}\vect{v} \quad \text{and} \quad  \alpha,
\label{eq:model:parameters:parameters}
\end{equation}
describe deformations of magnetic textures and, potentially, instabilities.

\subsection{Fundamental limit: the ferromagnetic instability}
\label{sec:model:ferromagnet}

An isolated skyrmion in a uniformly polarized background can only exist as long as the embedding phase is stable.
Therefore, the most fundamental upper limit for current-driven skyrmion motion is set by the instability of the polarized phase itself.~\cite{ Tserkovnyak2006,Bazaliy1998,Slonczewski1999,Fernandez-Rossier2004}
A linear order expansion of the LLG Eq.~\eqref{eq:dimensionless-LLG} around the ferromagnetic state with $\vect{m}_\mathrm{fm}=\hat{\vect{z}}$ yields the spectrum of spin waves which propagate as $\propto e^{i(\omega t - \vect{q}\cdot\vect{r})}$.
With this convention, the imaginary part of the spectrum reads
\begin{equation}
 \Im(\omega_\vect{q}^\pm) =  \frac{\alpha}{1 + \alpha^2}\left(  \pm \,\vect{q}\!\cdot\!\vect{u}+  \vect{q}^2 + 2\kappa \right).
\label{eq:model:ferromagnet:spectrum}
\end{equation}

For sufficiently small currents, the imaginary part of the spectrum is positive for all values of $\vect{q}$, $\Im(\omega_\vect{q}^\pm) > 0$, and hence the spin wave excitations decay exponentially to zero.
Above the critical drive 
\begin{equation}
\label{eq:criticaldrive}
|\vect{u}|=u_c^\mathrm{fm} = 2\,\sqrt{2\kappa},
\end{equation}
where $\Im(\omega_\vect{q}^\pm)$ may become negative, the excitations of the ferromagnetic phase grow exponentially, driving the ferromagnet unstable.
This sets an upper bound of possible drive parameters $u$ to investigate for STT-driven motion of any metastable state over the uniform background, including both domain walls and skyrmions.

\section{180$^\circ$ and 360$^\circ$ domain walls}
\label{sec:domainwalls}

\begin{figure}[t]
\includegraphics[width=\columnwidth]{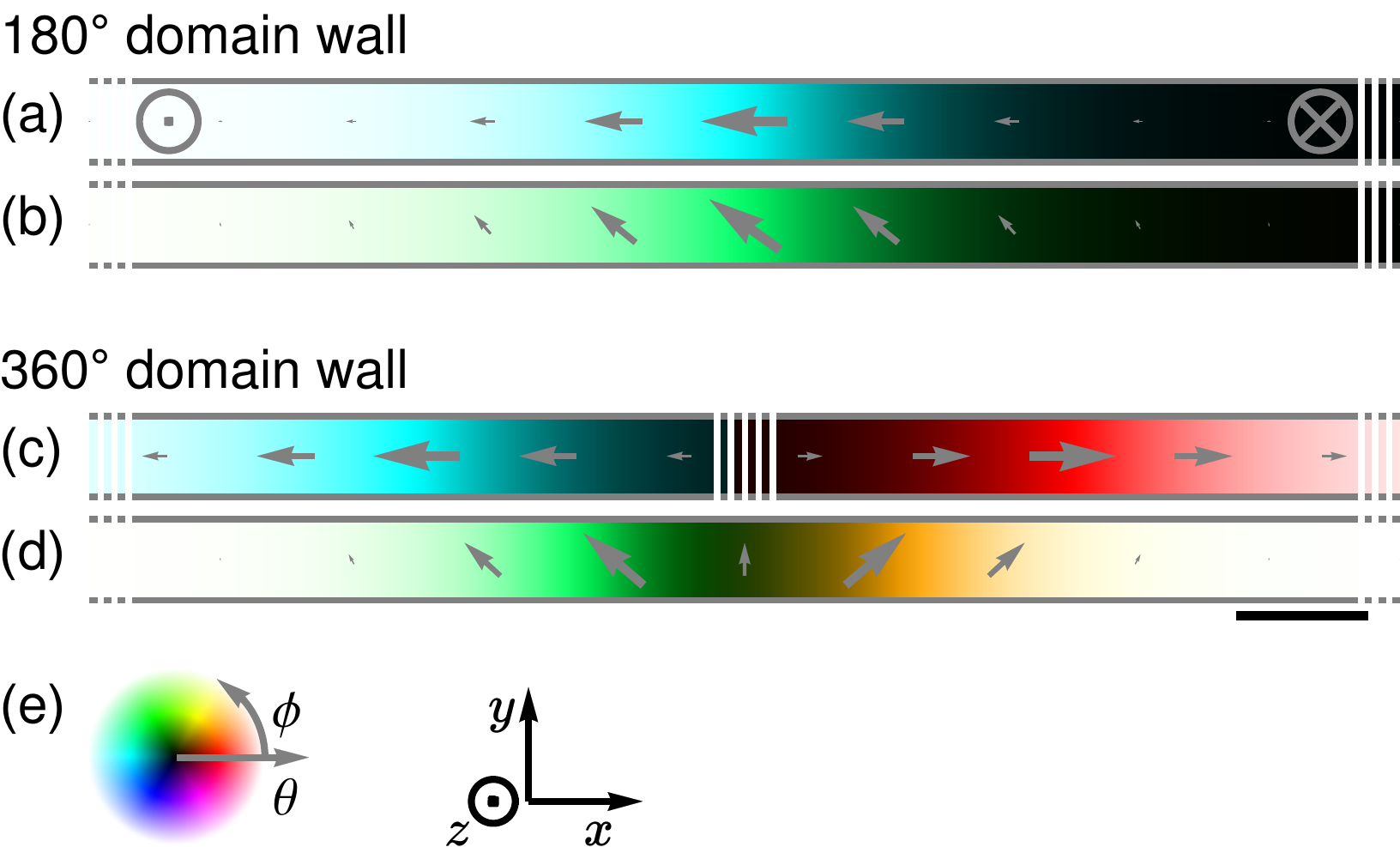} 
    \caption{  
    	Illustration of how spin currents lead to a canting of the magnetization out of the preferred plane of spin rotation set by the DM interaction.
        Magnetization of 180$^\circ$ and 360$^\circ$ domain walls in one spatial dimension in the absence (a,c) or presence (b,d) of STTs.
        The brightness encodes the azimuthal angle $\theta$ of the magnetization and the color the equatorial angle $\phi$ as indicated in (e).
        Gray arrows additionally illustrate the in-plane components.
        White vertical stripes denote infinite distances.
        The black line in (d) indicates one unit length.
        Parameters are $\kappa=1.3$ and $\vect{u}=\vect{0}$ (a,c) or $\vect{u}=0.95\,\hat{\vect{x}}= 0.29\, u_{c}^{\mathrm{fm}} \hat{\vect{x}}$ (b,d).
    }
    \label{fig:domainwalls:effect}
\end{figure}

In an effectively one-dimensional system, a 180$^\circ$ domain wall connects two oppositely out-of-plane polarized states, see Fig.~\ref{fig:domainwalls:effect}(a).
In a simple approximation, a skyrmion can be regarded as a closed 180$^\circ$ domain wall in two dimensions,~\cite{Romming2015} much as in early studies of magnetic bubbles.~\cite{Malozemoff1979}
Therefore, we first review and analyze the influence of STTs on domain walls in a simplified setup, neglecting magnetostatic interactions, to give first insights into the analogous effects for skyrmions, as a 360$^\circ$ domain wall is similar to a cross section through a skyrmion.

Without any applied STT, the magnetization across a chiral domain wall rotates in a plane which is set by the symmetry of the DM interaction.
For low current densities, but above an intrinsic pinning threshold, the STT leads to a steady translation and additional deformation of the domain wall.\cite{Li2004,He2006,Mougin2007} 
The dominant effect of the deformation is a canting of the magnetization out of the DM-plane, shown in Fig.~\ref{fig:domainwalls:effect}(b).
At a critical current, the magnetization in the center of the domain wall becomes perpendicular to its equilibrium position marking the onset of a Walker-esque breakdown.~\cite{Thiaville2005}$^{,}$~\footnote{Since it is due to STTs rather than a DC magnetic field, it is not the Walker breakdown in the original sense, however the $\beta$ term plays the role of an applied DC magnetic field.~\cite{Tatara2004}}
Above this critical current, the magnetization at the center of the wall precesses and the translational velocity of the wall oscillates in time.

This behavior can be understood by exploring the steady-state motion, Eq.~\eqref{eq:new-static-equation}.
Assuming that the domain wall moves at a constant speed $\vect{v}_{d}=\vect{v}_{\mathrm{dw}}$ without internal excitations, the solution to the Thiele Eq.~\eqref{eq:Thiele-equation} is $\vect{v}_{\mathrm{dw}}=(\beta/\alpha)\vect{v}$
and Eq.~\eqref{eq:new-static-equation} becomes
$\vect{0} = - \vect{m} \times \vect{B}^\mathrm{dw}_{\mathrm{eff}}$ with
\begin{equation}
\vect{B}^\mathrm{dw}_{\mathrm{eff}} = \vect{B}_{\mathrm{eff}} - \vect{m}\times( \vect{u} \cdot \nabla ) \vect{m}
\label{eq:domainwalls:LLG}
\end{equation}
where the influence of the applied current is absorbed into a modified effective magnetic field $\vect{B}^\mathrm{dw}_{\mathrm{eff}}$.
In terms of the set of reduced parameters, Eq.~\eqref{eq:model:parameters:parameters}, $\vect{B}_{\mathrm{eff}}$ depends only on $\kappa$.
Therefore, the deformation of the moving domain wall depends only on the effective coupling strengths $\kappa$ and the effective drive $u$.
For the up-to-down wall shown in Fig.~\ref{fig:domainwalls:effect}(a), Eq.~\eqref{eq:domainwalls:LLG} shows that the drive induces an extra effective magnetic field in the $\hat{\vect{y}}$-direction, explaining the deformation sketched in Fig.~\ref{fig:domainwalls:effect}(b).
Note that, for domain walls in systems with higher dimensions, the domain wall can further minimize its energy from the second term
in Eq.~\eqref{eq:domainwalls:LLG} by tilting its orientation.\cite{Muratov2017}

Without an applied current, an up-to-down domain wall as in Fig.~\ref{fig:domainwalls:effect}(a) can be transformed into its down-to-up counterpart via a time reversal transformation.
When a current is applied, this symmetry is broken by the modified $\vect{B}^\mathrm{dw}_{\mathrm{eff}}$ such that the magnetization in both types of domain walls twists along the same direction.
The current-induced asymmetry has strong implications on 360$^\circ$ domain walls, see Fig.~\ref{fig:domainwalls:effect}(c).
If stabilized by easy-axis anisotropy and without magnetic field, a 360$^\circ$ domain wall is unstable and decays into two separate 180$^\circ$ domain walls due to a repulsive force decaying exponentially with their distance $d_{\mathrm{dw}}$.\cite{Ghosh2017b}
When a current is applied, the STTs twist both 180$^\circ$ domain walls into the same direction and they become attractive.
The resulting bound state, see Fig.~\ref{fig:domainwalls:effect}(d), is, in fact, only a quasi-360$^\circ$ domain wall as the azimuthal angle does not cover the full 360$^\circ$, i.e., at no point $m_z=-1$.
Consequently, a sufficiently strong current can annihilate a pair of domain walls by smoothly unwinding it.\cite{Ghosh2017b}

We have calculated the equilibrium distance $d_{\mathrm{dw}}$ between two domain walls for different values of the coupling parameter
$\kappa$ and the effective drive $u$ from long-time numerical simulations of the LLG equation, Eq.~\eqref{eq:dimensionless-LLG}, in a moving frame of reference with the velocity $\vect{v}_\mathrm{dw}=(\beta/\alpha)\vect{v}$, for details of the numerical simulations see App.~\ref{sec:appendix:simulationdetails}.

\begin{figure}
    \includegraphics[width=0.336\textwidth]{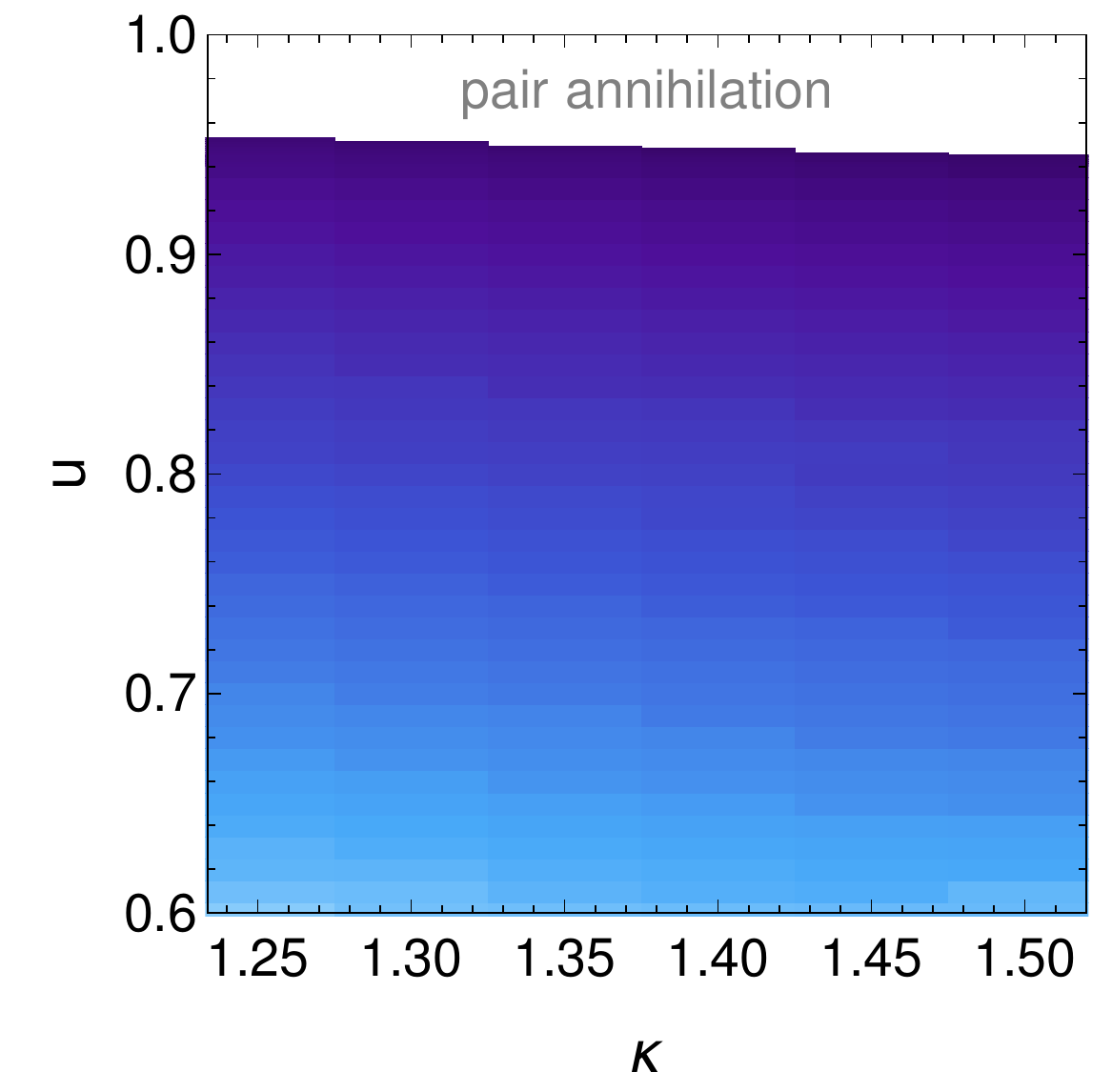}
    \hspace{0.002\textwidth}
    \includegraphics[width=0.038\textwidth]{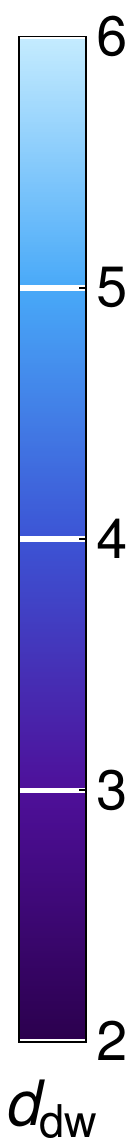}
    \caption{
    		Distance $d_\mathrm{dw}$ (encoded in color) between two domain walls in a current-induced bound state as function of coupling strength
		$\kappa$ and STT drive parameter $u$. It is defined as the distance between the two roots of $m_z(x)$.
    	In the white area, we do not find a stable bound pair.
	   }
    \label{fig:appendix:domainwalls}
\end{figure}

Our results for $u\geq0.6$ are summarized in Fig.~\eqref{fig:appendix:domainwalls}.
The critical current $u_c^\mathrm{bpdw}$ for the annihilation of the bound pair of domain walls decreases with increasing $\kappa$.

In the region considered, it can be well described by the linear fit $u_c^\mathrm{bpdw}(\kappa) \approx 0.9954 - 0.0336 \,\kappa$.
This domain wall pair annihilation occurs well below the ferromagnetic instability ($u_c^\mathrm{fm}= 2 \sqrt{2 \kappa} \gtrsim3$) or the Walker breakdown ($u_c^\mathrm{Walker}\gtrsim1.5$)\cite{Thiaville2005} at finite $d_{\mathrm{dw}}\gtrsim2.4$.

\section{Current-induced deformation of skyrmions}
\label{sec:skyrmion}

Similar to the 360$^\circ$ domain walls discussed in Sec.~\ref{sec:domainwalls}, skyrmions are also deformed by STTs. However,
unlike the domain wall bound pair, isolated skyrmions cannot be smoothly unwound by the antisymmetric effective magnetic field due to the STTs
due to their topological nature.
Moreover, the non-trivial winding number of a skyrmion gives rise to the skyrmion Hall effect, i.e., they do not move parallel to the driving current but, instead, along the direction $\vect{v}_{d}\!=\!\vect{v}_\mathrm{sky}$.

Assuming that the skyrmion moves as a rigid object, the time evolution of the magnetization is given by $\vect{m}(\vect{r}-\vect{v}_\mathrm{sky} t)$. In this case the skyrmion velocity $\vect{v}_\mathrm{sky}$
can be derived from the Thiele equation~\eqref{eq:Thiele-equation},
\begin{equation}
 \label{eq:skyrmion:thiele}
 \vect{v}_\mathrm{sky}
= \vect{v} + \frac{\alpha}{\mathcal{G}^2+\alpha^2\mathrm{det}(\mathcal{D})} \left( \vect{\mathcal{G}}\times(\mathcal{D}\vect{u}) - \alpha\, \mathrm{det}(\mathcal{D}) \vect{u} \right).
\end{equation}
Note that the dissipative matrix $\mathcal{D}$ depends on the profile of the skyrmion which potentially is deformed by the applied drive.
Furthermore, for the rigidly moving skyrmion Eq.~\eqref{eq:new-static-equation} becomes
\begin{align}
 \vect{0}
&=  - \vect{m} \times \left[ \vect{B}_\mathrm{eff} +
\alpha \left(\vect{v}_\mathrm{DL}\cdot\nabla\right) \vect{m} + \vect{m}\times \left(\vect{v}_\mathrm{FL}\cdot\nabla\right) \vect{m}\right]
  \label{eq:skyrmion:movingLLG}
\end{align}
where the two current-induced effective fields are mediated by $\vect{v}_\mathrm{FL} = \vect{v}_\mathrm{sky} - \vect{v}$ and $\vect{v}_\mathrm{DL} = \vect{v}_\mathrm{FL} +\vect{u}$.
The term proportional to $\vect{v}_\mathrm{DL}$ is dissipative in the sense that it is the only torque that is odd under time reversal,~\footnote{Under $t\mapsto -t$, the quantities transform as: $\vect{v}_{i}\mapsto-\vect{v}_{i}$, $\vect{m}\mapsto -\vect{m}$, $\vect{B}_{\mathrm{eff}}\mapsto - \vect{B}_{\mathrm{eff}}$.} and meanwhile the $\vect{v}_\mathrm{FL}$ term is a reactive torque due to the applied current.

We point out here that in the Galilean invariant case, i.e. $\alpha = \beta$, the drive $\vec{u}$ vanishes and so do the effective current-induced torques in Eq.~\eqref{eq:skyrmion:movingLLG}, such that the skyrmion shape is unmodified by the current, as expected.
However, for any finite $\vec{u}>\vect{0}$, i.e.\ $\alpha \neq \beta$, both $\vect{v}_\mathrm{FL}$ and $\vect{v}_\mathrm{DL}$ are always simultaneously non-zero (see App.~\ref{app:v1-v2-analysis}) and lead to a deformation of the skyrmion.
Moreover, they both depend on the dissipative matrix $\mathcal{D}$ which  itself depends on the deformations due to the applied current. Therefore, Eqs.~\eqref{eq:skyrmion:thiele} and \eqref{eq:skyrmion:movingLLG} must be solved self-consistently.

As a quantitative measure for the deformation of the skyrmion, we introduce two order parameters: (i) the \emph{distortion} $\delta$ which quantifies how close the skyrmion is to a circular shape and (ii) the \emph{axis of distortion} $\vec d$ which corresponds to the principle axis of the deformation.
Both these quantities are extracted from the nontrivial $2\times2$ block of the dissipation matrix $\mathcal D$ which is symmetric and can therefore always be diagonalized.
With the normalized eigenvectors as $\vec{\lambda}^\pm$ and their corresponding eigenvalues $\lambda^\mathrm{+} > \lambda^\mathrm{-} > 0$ we define
\begin{equation}
 \delta \equiv \frac{ \lambda^\mathrm{+} - \lambda^\mathrm{-} }{ \lambda^\mathrm{+} } \quad \text{and} \quad \vec{d} = \vec{\lambda}^\mathrm{-} .
  \label{eq:skyrmion:distortion}
\end{equation}
With this convention, $\delta=0$ corresponds to a perfectly circular skyrmion, while $\delta=1$ is an arbitrarily elongated skyrmion.
We choose the sign of $\vect{d}$ such that the gyrotropic vector $\vec{\mathcal{G}}$ ($\sim$ skyrmion center orientation), the drive $\vec u$, and the direction of distortion $\vec d$ obey a right handed rule: $(\vec{\mathcal{G}} \times \vec u)\cdot  \vect d >0$.

The deformations and instabilities of skyrmions, are discussed in the following based on our numerical results and analytical approximations.
For an extended discussion on the simulation details see App.~\ref{sec:appendix:simulationdetails}.

\subsection{Phase diagrams for current-driven skyrmions}
\label{sec:skyrmion:phasediagrams}

\begin{figure}
    \includegraphics[width=0.072\textwidth]{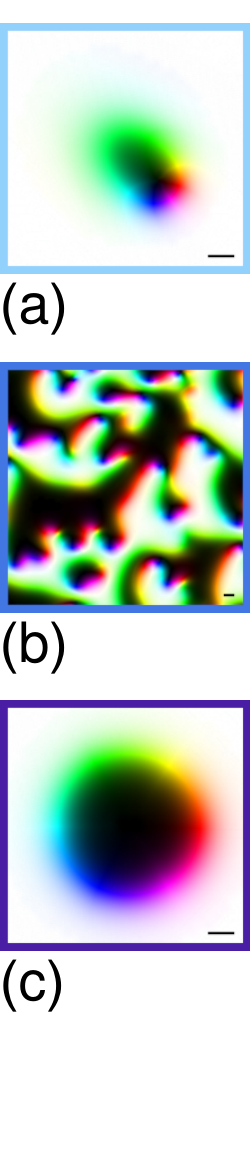}
   \hspace{0.002\textwidth}   
    \includegraphics[width=0.336\textwidth]{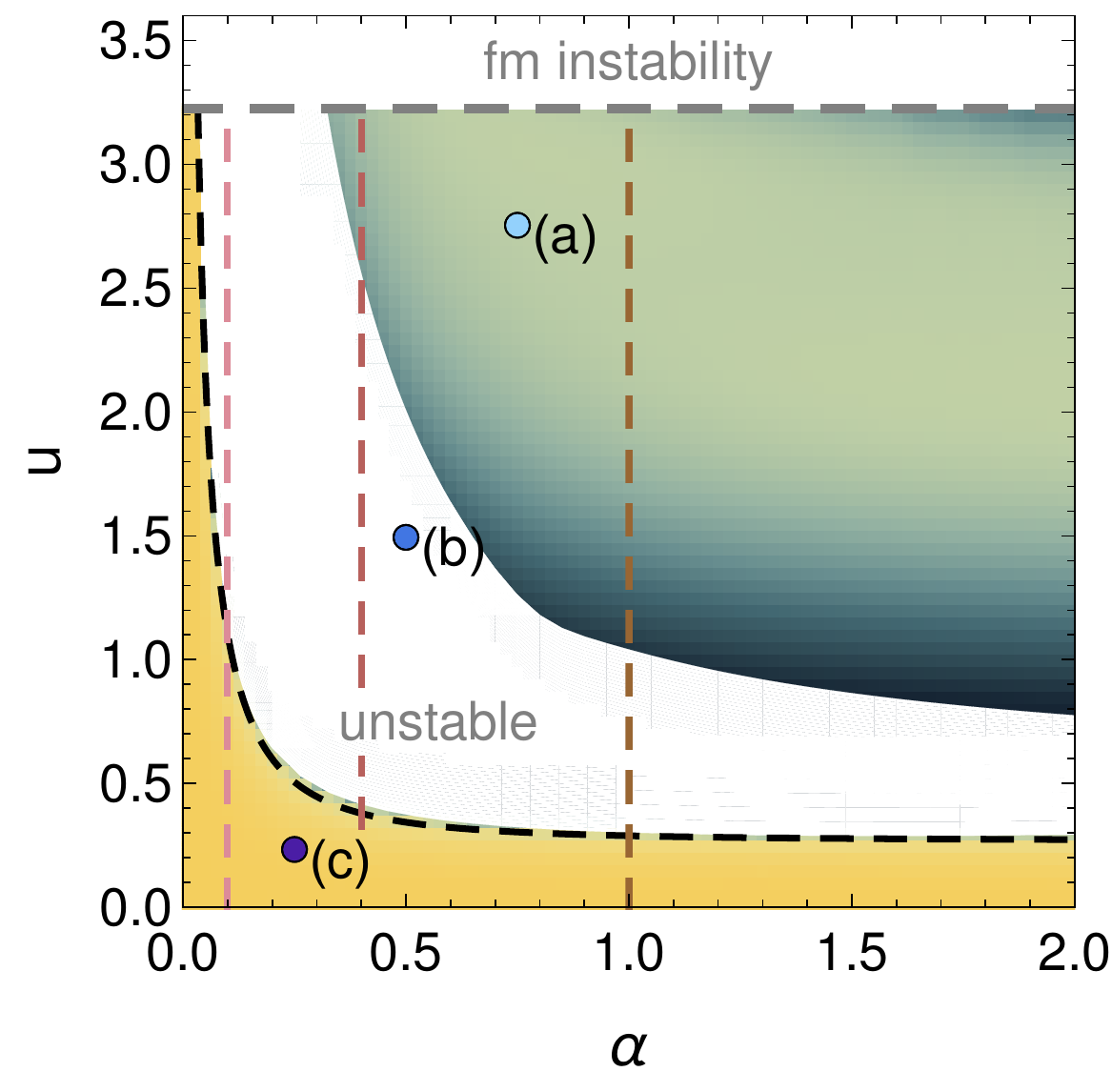}
   \hspace{0.002\textwidth}
    \includegraphics[width=0.053\textwidth]{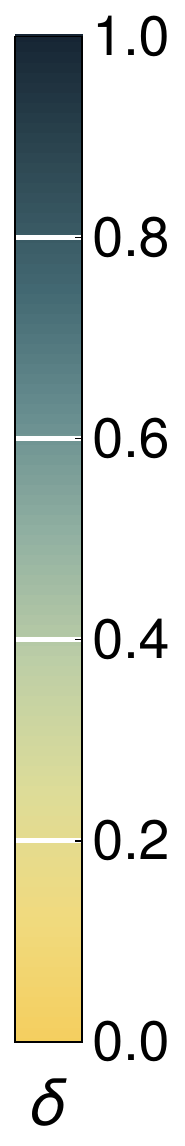}
    \caption{
    	Phase diagram for a skyrmion as function of Gilbert damping $\alpha$ and drive parameter $u$ for anisotropy $\kappa=1.3$.
        The color gradient (yellow to dark green) encodes the distortion $\delta$, see colorbar.
        In white areas, no steady skyrmion solution exists.
        The dashed gray horizontal line indicates the ferromagnetic instability, see Eq.~\eqref{eq:criticaldrive} in
        Sec.~\ref{sec:model:ferromagnet}.
        The black dashed line corresponds to the analytical expression for the onset of the instability region, see 
        Eq.~\eqref{eq:instabilitycondition} in Sec.~\ref{sec:skyrmion:elliptical}. 
        Vertical dashed lines mark cuts at constant $\alpha$, corresponding to the cuts in the phase diagrams in 
        Fig.~\ref{fig:skyrmion:phasediagrams:constAlpha}.
        Blue dots indicate the position of the corresponding real-space magnetization textures shown in the left panels: (a) a shooting star-like distorted skyrmion, (b) no stable skyrmion, and (c) a rather circular skyrmion.
        }
    \label{fig:skyrmion:phasediagrams:constK}
\end{figure}

\begin{figure*}  
    \includegraphics[width=0.336\textwidth]{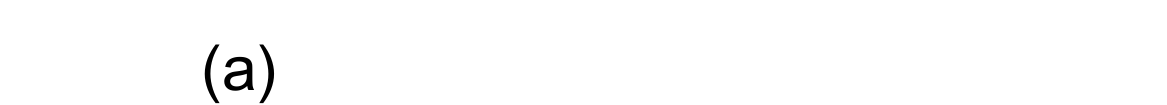}
    \hspace{0.001\textwidth}
    \includegraphics[width=0.285\textwidth]{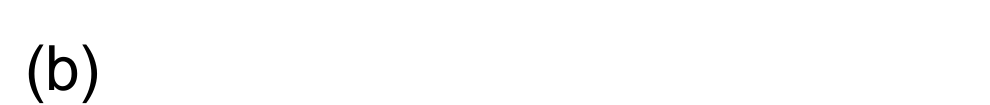}
    \hspace{0.0000\textwidth}
    \includegraphics[width=0.285\textwidth]{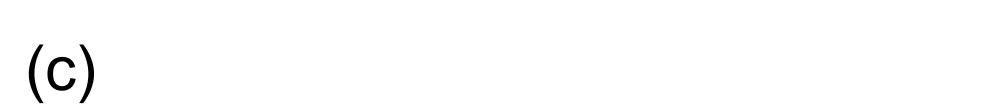}
    \hspace{0.002\textwidth}
    \hspace{0.053\textwidth}
    
    \includegraphics[width=0.336\textwidth]{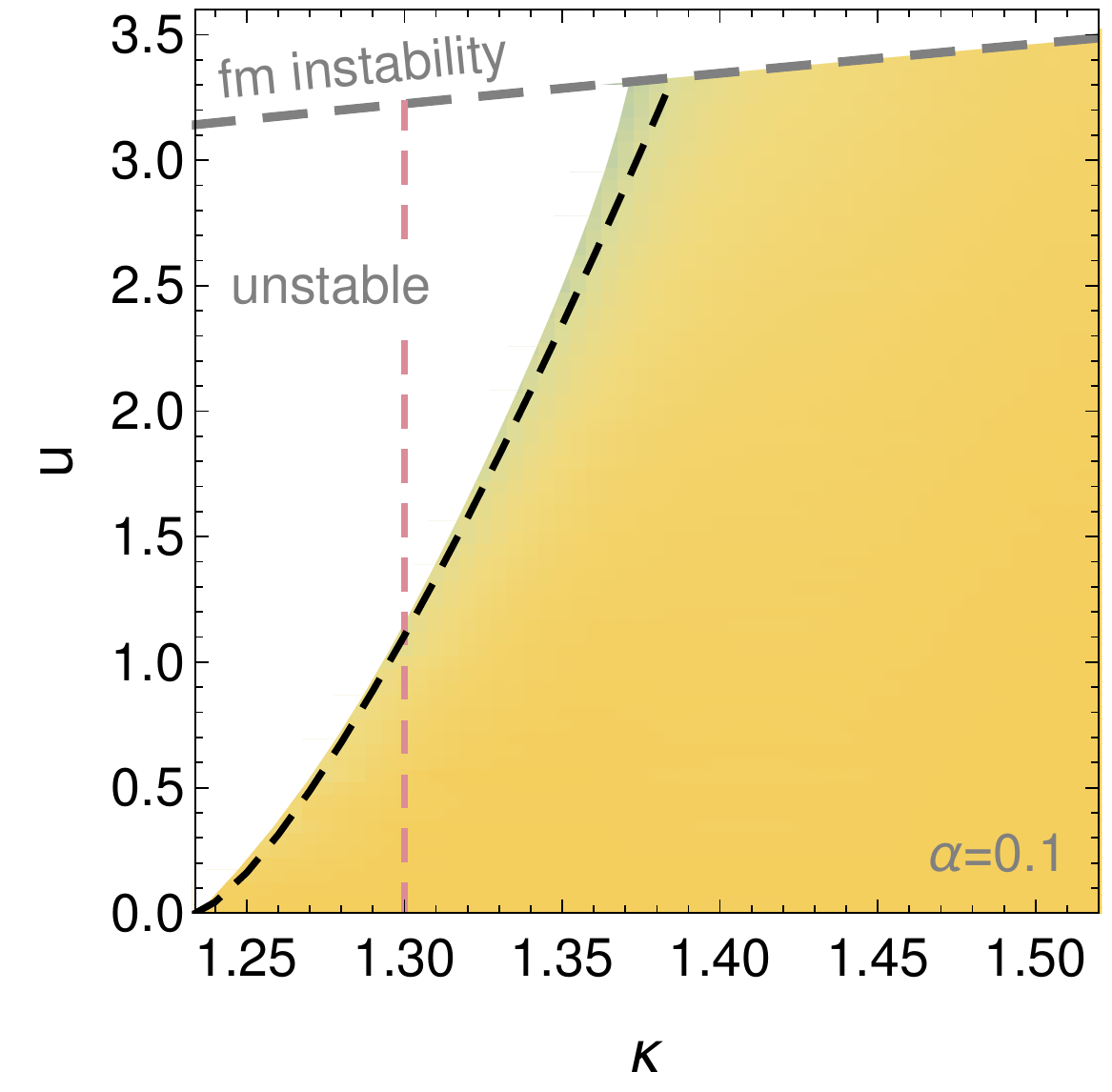}
    \hspace{0.001\textwidth}
    \includegraphics[width=0.285\textwidth]{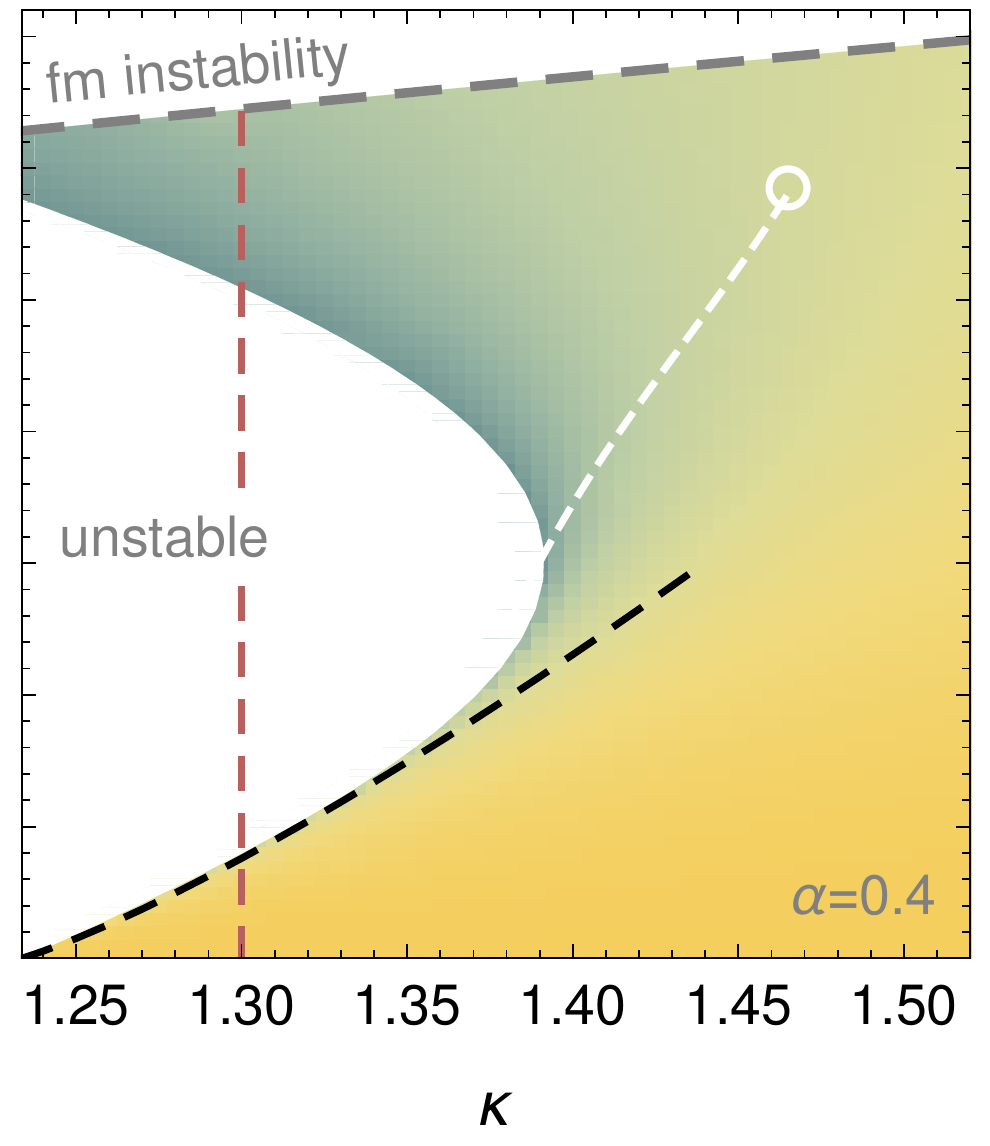}
    \hspace{0.0000\textwidth}
    \includegraphics[width=0.285\textwidth]{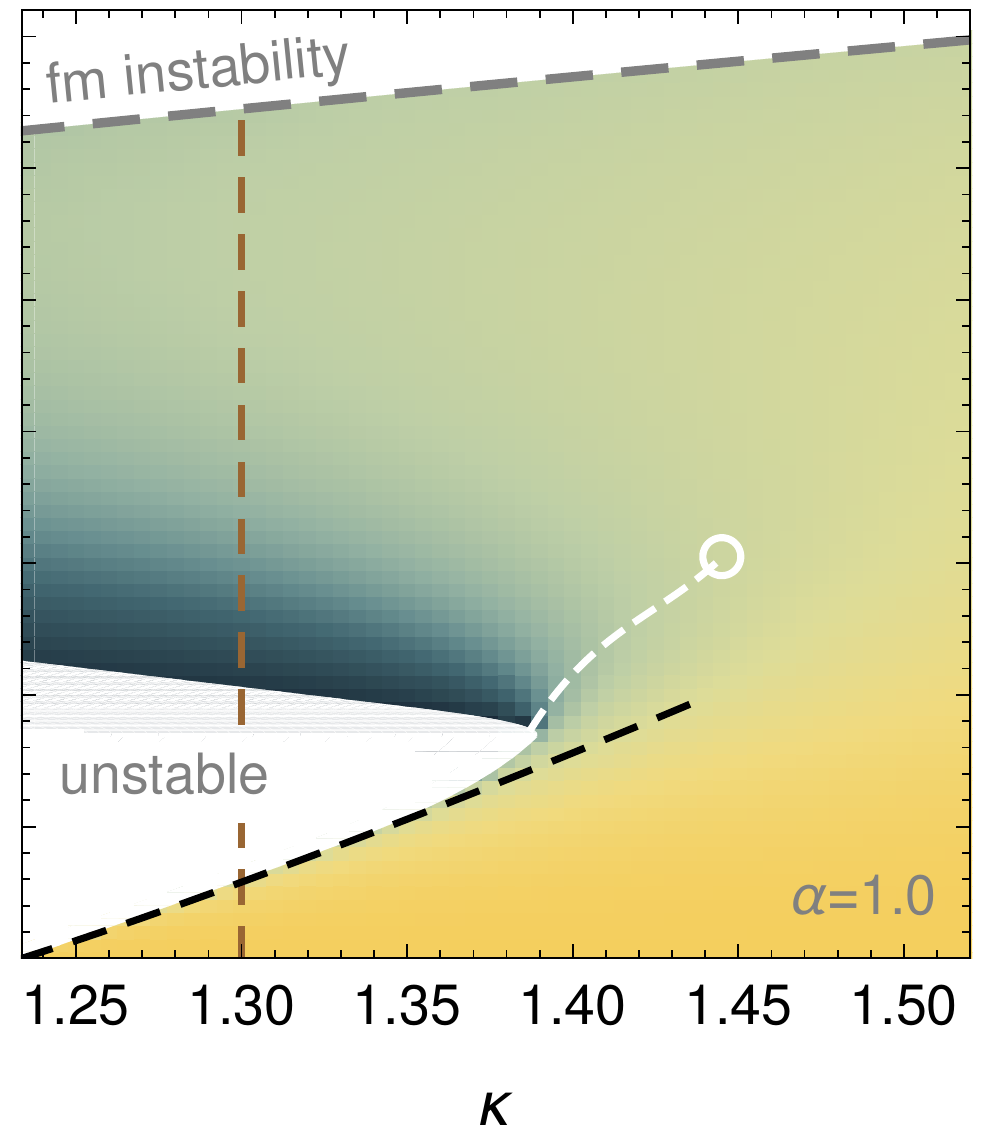}
    \hspace{0.002\textwidth}
    \includegraphics[width=0.053\textwidth]{fig3c.pdf}
    \caption{
        Distortion $\delta$ of a  skyrmion as function of anisotropy $\kappa$ and drive parameter $u$ for different values of the Gilbert
        damping $\alpha=0.1, 0.4, 1.0$.
        As in Fig.~\ref{fig:skyrmion:phasediagrams:constK} $\delta=0$ (yellow) corresponds to an unperturbed (circular) skyrmion
        and $\delta=1$ (dark green) to an infinitely elongated skyrmion.
        White areas indicate regimes for which no steady skyrmion solution exists.
        The black dashed line corresponds to the analytical expression for the onset of the instability region, see 
        Eq.~\eqref{eq:instabilitycondition} in Sec.~\ref{sec:skyrmion:elliptical}.         
        The dashed gray line indicates the ferromagnetic instability, see Eq.~\eqref{eq:criticaldrive} in Sec.~\ref{sec:model:ferromagnet}.
        The vertical dashed lines mark cuts at constant $\kappa=1.3$, corresponding to the cuts in the phase diagram in
        Fig.~\ref{fig:skyrmion:phasediagrams:constK}.
        The white dashed line indicates a local maximum of $\delta(u)$ for fixed $\kappa$ which smoothly flattens out for larger $\kappa$ and
        finally disappears (white circle). For detailed analysis, see App.~\ref{app:phasediagramdetails}.
        }
    \label{fig:skyrmion:phasediagrams:constAlpha}
\end{figure*}

In this part, we systematically investigate the deformation parameter $\delta$ of isolated skyrmions under the influence of STTs  as a function of the three parameters the anisotropy coupling $\kappa$, drive parameter $\vect{u}$ and Gilbert damping, $\alpha$; i.e.\ $\delta \equiv \delta(\kappa,u,\alpha)$.
In our study, we consider only regimes where the ferromagnet is the ground state, i.e.\ $\kappa > \kappa_{c}= \pi^2/8 \sim 1.23$, and cover all reduced drives $u$ below the ferromagnetic instability, $u< u_c^\mathrm{fm}=2\sqrt{2\kappa}$.

We start by investigating the dependence of the deformation parameter $\delta$ in terms of $\vect{u}$ and $\alpha$ for $\kappa = 1.3 \approx 1.05\, \kappa_{c}$, where strong deformations are observed. Our numerical results are shown in Fig.~\ref{fig:skyrmion:phasediagrams:constK}.
We notice the existence of three main areas:
1) The white area corresponds to the region where no steady skyrmion solution is possible, i.e.\ skyrmions are unstable;
2) the mainly yellow area below the instability shows the region where the skyrmions remain mostly circular, $\delta \approx 0$, see Sec.~\ref{sec:skyrmion:deformation};
and 3) the area above the instability, where $\delta \gtrsim 0.4$.

Among the interesting features of this phase diagram is that, as we increase the drive parameter $u$ from the region with almost circular skyrmions towards the unstable region, there is an \textit{elliptical instability}, see Sec.~\ref{sec:skyrmion:elliptical}, where the skyrmions elongate and eventually collapse or keep elongating indefinitely, see Fig.~\ref{fig:skyrmion:elliptical:realspace}.
The stability, however, is recovered for large $\alpha$ upon increasing the drive, where the skyrmion relaxes to a steady solution with an asymmetric shape that resembles a shooting star, see Sec.~\ref{sec:skyrmion:shootingstars}.

In Fig.~\ref{fig:skyrmion:phasediagrams:constAlpha} we show the dependence of the deformation parameter $\delta$ for a skyrmion in terms of the  drive $u$ and the anisotropy strength $\kappa$ ($\sim$ inverse skyrmion size) for different damping values $\alpha$.
While for small damping parameters we find only the transition from mainly symmetric skyrmions to the unstable region via the elliptical instability, the behavior for large $\alpha$ is more complex.
For low $\kappa$ and low $u$ the situation is similar, i.e.\ upon increasing the drive the circular skyrmion distends until reaching the unstable region. 
Continuing on to even higher $u$, past this apparent instability of isolated steady-state traveling wave solutions, we find the regime where the steady state resembles a shooting star.
 Above a certain coupling parameter strength $\kappa$, radially symmetric skyrmions smoothly transform into shooting-star-like skyrmions upon increasing the drive. 
 For detailed plots of the distortion as a function of drive for various $\kappa$ we refer to App.~\ref{app:phasediagramdetails}.

In the following sections we will present in detail the possible steady skyrmion configurations obtained from our numerics, discuss their properties and explain the elliptical instability.
To obtain a physical understanding of our numerical results, we will discuss how torques act on different magnetic configurations.

\begin{figure}
    \includegraphics[width=0.235\textwidth]{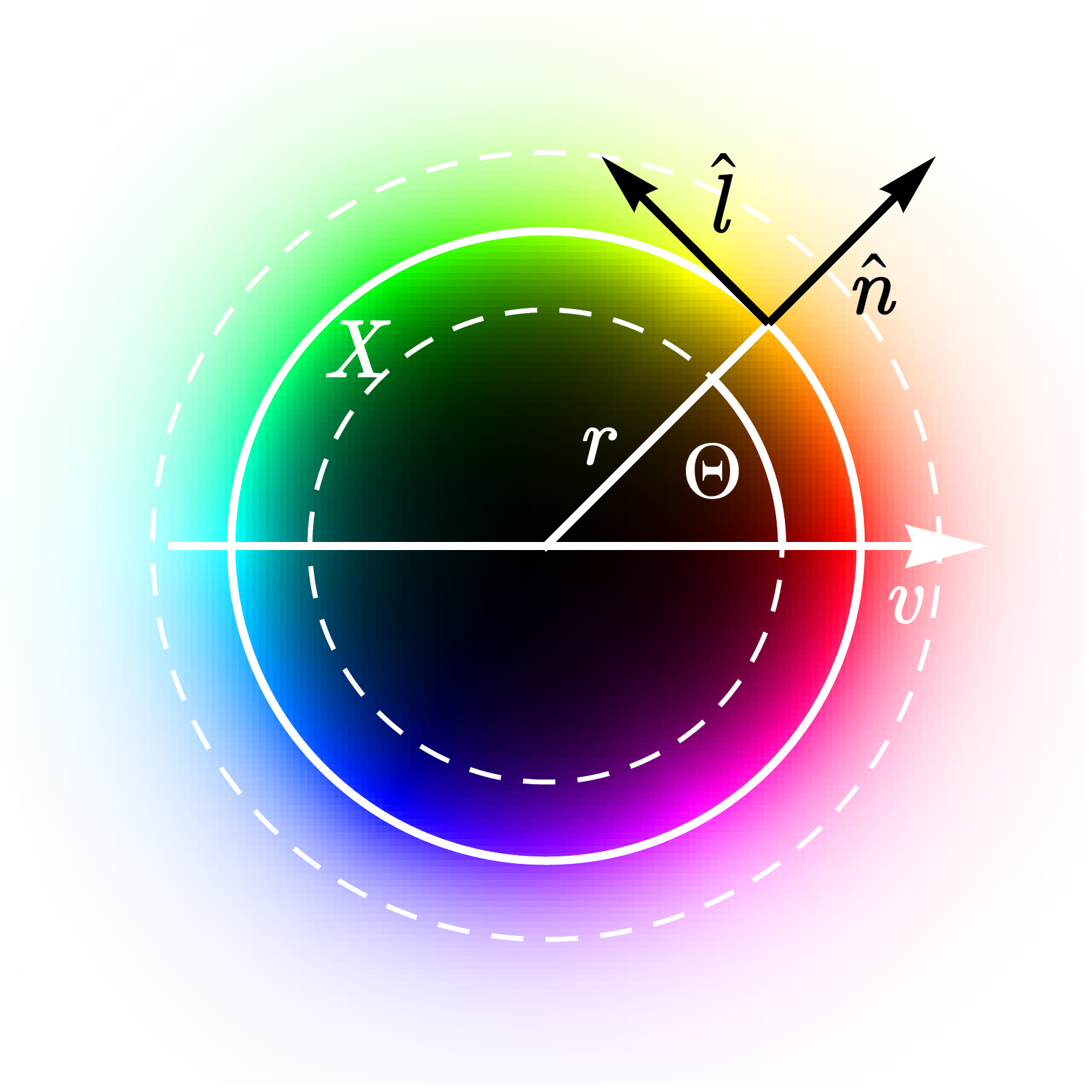}
    \includegraphics[width=0.235\textwidth]{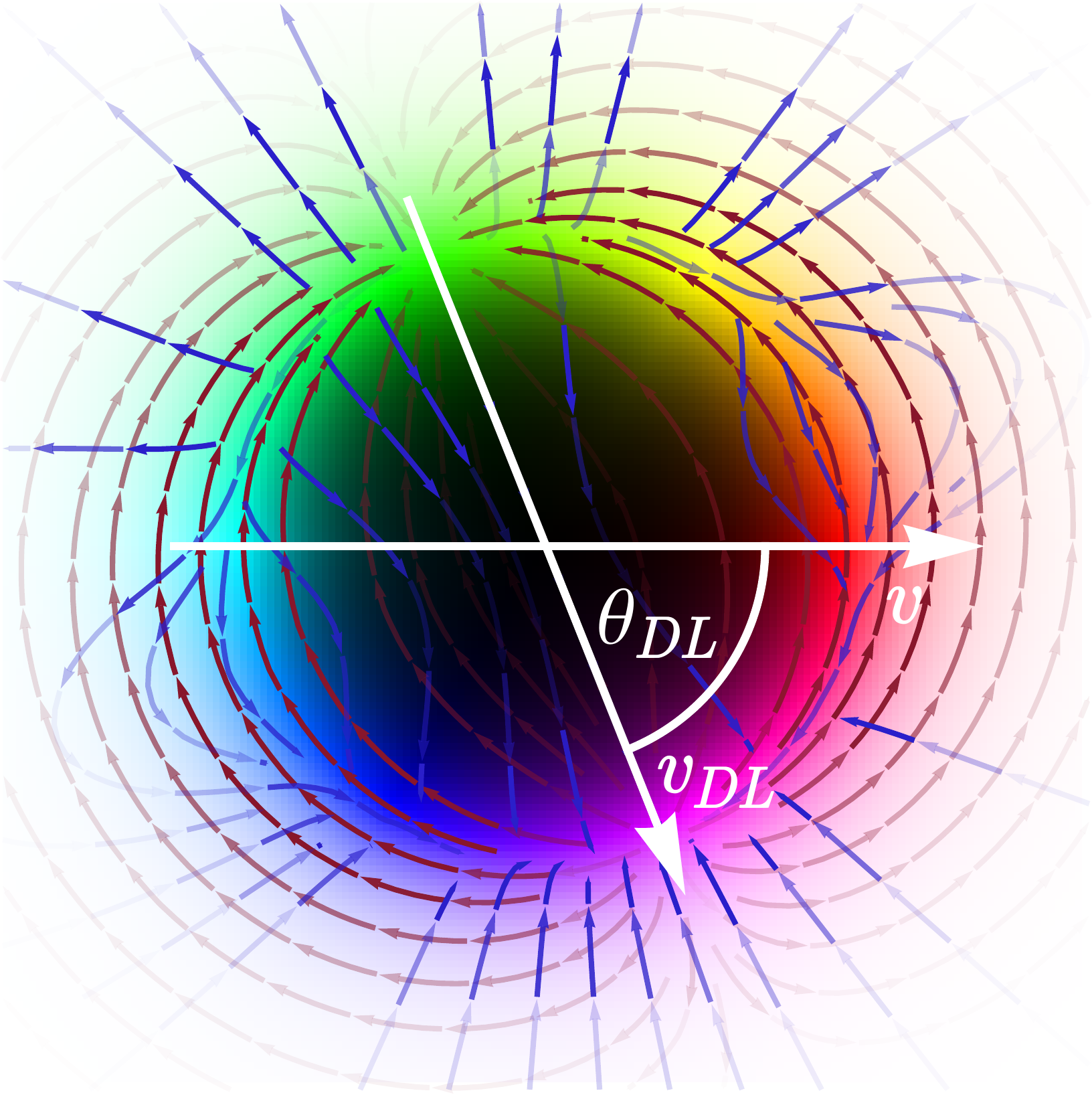}
   \caption{
        Magnified version of the slightly deformed skyrmion shown in Fig.~\ref{fig:skyrmion:phasediagrams:constK}, $\delta\approx0$. 
        The horizontal white arrow represents the direction of the current $\vect v$.
        Left panel: Coordinate system in the vicinity of the skyrmion boundary. 
        The curve where the magnetization is in-plane is given by $\vect{X}(l)$ (solid circle) with the corresponding locally defined tangential vector $\hat{\vect l}$ and normal vector $\hat{\vect n}$.
        Right panel: Field lines depict the damping-like (blue) and field-like (red) parts of the effective magnetic field, Eqs.~\eqref{eq:BDL} and \eqref{eq:BFL}. 
        For a mirror symmetric skyrmion, as shown, the main axis of distortion $\vect d$ corresponds to the direction of $\vect{v}_{\mathrm{DL}}$. 
       }
   \label{fig:skyrmionboundary}
\end{figure}

For this let us define the border of the skyrmion as the curve along which the magnetization is in-plane, $\vect{X}(l)$, where $l$ is the arc-length,\cite{Rodrigues2018, Kravchuk2019, Litzius2020} see Fig.~\ref{fig:skyrmionboundary}.
 We define a basis of the normal and tangential unit vectors as $\hat{\vect{l}} = \partial_{l}\vect{X}, \,\textrm{and}\, \hat{\vect{n}} = \hat{\vect{l}}\times\hat{\vect{z}}$. With this definition, the local radius of curvature $r(l)$ is $r(l) = (\hat{\vect{l}}.\partial_{l}\hat{\vect{n}})^{-1}$ and a position $\vect x$ in the vicinity of the border can be parameterized as
$\vect{x} = \vect{X}(l) + n\hat{\vect{n}}(l)$,
with $n\ll r(l)$. The magnetization in this region can then be defined in spherical coordinates as
\begin{equation}\label{eq:magnetizationborder}
\vect{m} = \sin\theta\cos\phi\, \hat{\vect{n}} + \sin\theta\sin\phi \, \hat{\vect{l}} + \cos\theta \, \hat{\vect{z}}.
\end{equation}
Substituting this ansatz into Eq.~\eqref{eq:skyrmion:movingLLG}, using that $\theta\equiv\theta(n,l)$, $\hat{\vect{l}} \equiv \hat{\vect{l}}(l)$, $\hat{\vect{n}} \equiv \hat{\vect{n}}(l)$,
and assuming $\phi(n,l)\equiv\phi(l)$ is constant along the normal direction.\footnote{Note that $\phi(n,l)\equiv\phi(l)$  is in particular true for radially symmetric skyrmions.~\cite{Rohart2013,Kravchuk2018,McKeever2019,Rodrigues2018,Kravchuk2019} 
In this case, in the absence of current, Eqs.~\eqref{eq:skyrmionradialprofile} a) and b) correspond to the big radius limit of the profile studied in Ref.~\onlinecite{Leonov2016a}.},
we obtain the effective equations defining the profile of the magnetization in the vicinity of the steady skyrmion boundary in the presence of STTs
\begin{subequations}\label{eq:skyrmionradialprofile}
\begin{align}
\label{eq:skyrmionradialprofile1}
0= & \ \partial_{n}^2\theta + \partial_{l}^2\theta + 2\Omega\sin^2\theta\cos\phi\\
& \ - \frac{\sin2\theta}{2}\left(2\kappa + \Omega^2 + (\partial_{l}\theta)^2 - 2\partial_{l}\theta\sin\phi\right) \notag\\
& \ +\alpha v_{\mathrm{DL}}\left(\cos(\theta_{\mathrm{DL}} - \Theta)\partial_{n}\theta + \sin(\theta_{\mathrm{DL}} - \Theta)\cos^2\theta\partial_{l}\theta\right)\notag\\
& \ - v_{\mathrm{FL}}\sin(\theta_{\mathrm{FL}} - \Theta)\Omega\sin\theta,\notag\\[2mm]
\label{eq:skyrmionradialprofile2}
0= &2 \partial_{n}\theta\sin\theta\sin\phi +2\partial_{l}\theta\cos\theta\Omega + \sin\theta\partial_{l}\Omega\\
 &\ +\alpha v_{\mathrm{DL}}\sin(\theta_{\mathrm{DL}} - \Theta)\sin\theta\Omega\notag\\
 &\ + v_{\mathrm{FL}}\left(\cos(\theta_{\mathrm{FL}} - \Theta)\partial_{n}\theta + \sin(\theta_{\mathrm{FL}} - \Theta)\cos^2\theta\partial_{l}\theta \right),\notag
 \end{align}
\end{subequations}
where $\Omega(l) \equiv \partial_{l}\phi + 1/r(l)$. The angles $\Theta(l), \theta_{\mathrm{DL}}$ and $\theta_{\mathrm{FL}}$ are the angles between $\hat{\vect n}$, $\vect{v}_{\mathrm{DL}}$ and $\vect{v}_{\mathrm{FL}}$ with respect to $\vect{u}$, respectively, see Fig.~\ref{fig:skyrmionboundary}.

Notice that these equations are analogous to the usual equations for domain wall dynamics in one dimension, \cite{Slonczewski1973, Thiaville2002, Tretiakov2010}
with two main differences: First, for one-dimensional domain walls the local radius of curvature is zero and the function $\Omega$ converts to $\Omega \rightarrow \partial_{n}\phi$. Second, for a skyrmion both damping and field-like terms contribute to a deformation of the shape, while for a domain wall only the field-like term contributes, see Eq.~\eqref{eq:domainwalls:LLG}.
The reason for the latter is that while for domain walls  $v_{\mathrm{DL}}$ can be set to zero independently from $v_{\mathrm{FL}}$, this is not possible for skyrmions, see App.~\ref{app:v1-v2-analysis}.

In general, Eq.~\eqref{eq:skyrmionradialprofile1} is associated to the profile of the magnetization along the normal direction. For a circular skyrmion Eq.~\eqref{eq:skyrmionradialprofile2} is trivial and just defines the global azimuthal angle. Whenever we consider deformations of the skyrmion these two equations are coupled and must be solved self-consistently.

While the gyrotropic vector $\vec{\mathcal{G}}$ is associated to the topological property of the skyrmion, and is, therefore, invariant under smooth transformations, the dissipative matrix $\mathcal{D}$ depends explicitly on the shape of the skyrmion. If we substitute the ansatz \eqref{eq:magnetizationborder} of the skyrmion border into the definition of $\mathcal{D}$ we obtain the following components
\begin{subequations}\label{eq:DMatrixDW}
\begin{align}
\mathcal{D}_{xx} &=\frac{1}{2} \int dl dn \bigl(\Delta_{n, \Omega}^{+}+ \Delta_{n, \Omega}^{-} \cos2\Theta \bigr), \\
\mathcal{D}_{yy} &=\frac{1}{2} \int dl dn \bigl(\Delta_{n, \Omega}^{+} - \Delta_{n, \Omega}^{-} \cos2\Theta \bigr), \\
\mathcal{D}_{xy} &= \frac{1}{2}\int dl dn \bigl(\Delta_{n, \Omega}^{-} \sin 2\Theta \bigr),
\end{align}
\end{subequations}
where 
we have introduced $\Delta_{n, \Omega}^{\pm} = \partial_{n}\theta^2 \pm \Omega^2 \sin ^2\theta$, 
and the integrals are performed over the skyrmion border.
Notice that an axially symmetric skyrmion is described by $\Theta = \psi$ where $\psi$ is the polar angle, $\Omega = 1/R$ where $R$ is the skyrmion radius, and $\theta(n,l) = \theta(n)$. It therefore follows that $\mathcal{D}_{xx} = \mathcal{D}_{yy}$ and $\mathcal{D}_{xy}=0$, as expected.

We also would like to highlight another special case, namely when the skyrmion is mirror symmetric with respect to the axis given by $\vect{v}_{\mathrm{DL}}$. 
Then 
i) the two effective velocities are perpendicular to each other, $\vect{v}_{\mathrm{DL}} \perp \vect{v}_{\mathrm{FL}}$, 
ii) they correspond to the eigenvectors of the dissipative matrix $\mathcal{D}$, 
and iii) $\vect{v}_{\mathrm{DL}}$ is the axis of distortion, i.e., $\vect{d} \parallel \vect{v}_{\mathrm{DL}} $, see App.~\ref{app:symmetry}. 

\subsection{Skyrmion deformations at low drives}
\label{sec:skyrmion:deformation}

In the absence of a driving current, the shape of a skyrmion is axially symmetric.
Then the corresponding dissipative matrix reduces to a simple diagonal form,~\footnote{Note: as vectors in the Thiele equation~\eqref{eq:Thiele-equation} point in the $(x,y)$--plane, the $\mathcal{D}_{zz}$ matrix element is irrelevant.} $\mathcal{D}=\mathbb{1} \mathcal{D}_s$, and there is no distortion, i.e.\ $\delta = 0$, see Eq.~\eqref{eq:skyrmion:distortion}. 
When a driving current is applied, the rotational symmetry is broken and the skyrmion is deformed along an axis which finally is determined by $\vect{v}_{\mathrm{DL}}$ and $\vect{v}_{\mathrm{FL}}$, see Eq.~\eqref{eq:skyrmion:movingLLG}.

For small driving parameters, we can perform a linear expansion in $\vect u$. 
While the deformation itself scales linearly with $\vect u$,  the dissipative matrix does not have corrections to linear order in $\vect u$, $\mathcal{D}=\mathbb{1} \mathcal{D}_s + \mathcal{O}(u^2)$. 
Thus, at low drives, the skyrmion remains mirror symmetric with respect to the axis of the deformation.
A more detailed analysis of Eq.~\eqref{eq:skyrmionradialprofile} to linear order in the drive is given in App.~\ref{app:lowdrive}.
It follows that the skyrmion is mirror symmetric with respect to $\vect{v}_{\textrm{DL}}$, thus 
$\vect{d} \parallel \vect{v}_{\mathrm{DL}}$ and the first non-zero contribution to the distortion is given by
\begin{equation}
\delta \propto \frac{1}{R^2} \left((\alpha v_{\mathrm{DL}} - v_{\mathrm{FL}})^2 - c(\kappa/\kappa_c)\, \alpha\, v_{\mathrm{DL}}v_{\mathrm{FL}}\right).
\end{equation} 
Here, $R \approx 1/\left(2\sqrt{\kappa - \sqrt{\kappa \kappa_{c}}}\right)$ is the approximate radius of a circular skyrmion~\cite{Rohart2013,Kravchuk2018,McKeever2019} and $c(\kappa/\kappa_{c})$ depends on the details of the skyrmion solution.
For $\kappa\rightarrow \kappa_{c}$ one can expand $c$ in terms of $\kappa$ such that to leading order we obtain a constant. 

\begin{figure}
    \includegraphics[width=0.47\textwidth]{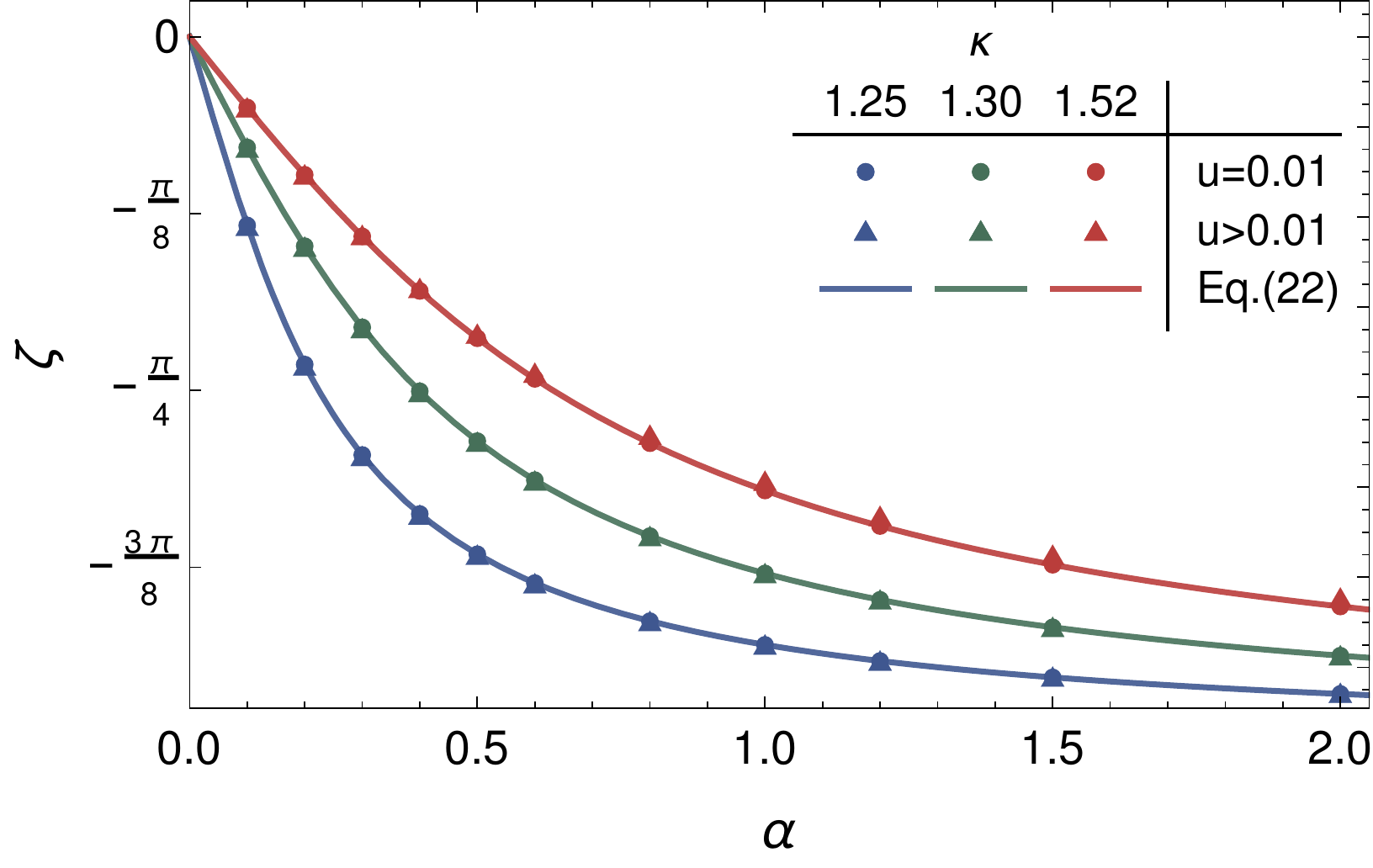}
    \caption{
        Angle $\zeta$ between the axis of distortion $\vect{d}$ and the effective drive $\vect{u}$ as a function of the Gilbert damping $\alpha$ for anisotropies $\kappa=1.25$, $1.30$, and $1.52$.
        Solid lines indicate the analytic approximation for $u=0$, see Eq.\eqref{eq:skyrmion:theta}.
        Numerical results are shown for a very low drive $u=0.01$ (dots) and an elevated drive $u=0.07$, $0.25$, and $1.0$, respectively (triangles). 
        They are in agreement with our analytical prediction and are, in particular, independent of the drive.
       }
    \label{fig:skyrmion:deformation:axis}
\end{figure}

Furthermore, for small drive parameters, the angle $\zeta$ between the deformation axis $\vec d$ and the drive $\vec u$ is given by
\begin{align}
 \zeta =& \angle(\vec{u},\vec{d}) = \arctan\left(  \frac{      4\pi \mathcal{Q} \, \alpha \, \vec{u}\cdot(\mathcal{D}\vec{u})        }{         \mathcal{G}^2 u^2 + \alpha \, \vec{u}\cdot (\vec{\mathcal{G}}\!\times\!(\mathcal{D}\vec u))         } \right)\notag\\  \approx& \arctan \left(-\frac{\alpha \mathcal{D}_s}{4\pi}\right),
 \label{eq:skyrmion:theta}
\end{align}
where we have used $\mathcal{D}\vec{u} \approx \mathcal{D}_s \vec{u}$ up to first order in the drive. 
We show a comparison of this approximate expression, being independent of the drive parameter, and the results of numerical simulations in Fig.~\ref{fig:skyrmion:deformation:axis}.
Even for large STTs with a drive parameter of $u=1.0$, the distortion $\delta=0.07$ is rather small, and the numerical results agree very well with the predicted expression in Eq.~\eqref{eq:skyrmion:theta}. Moreover, as expected, the axis of distortion $\vect{d}$ changes with the effective anisotropy parameter $\kappa$, since $\mathcal{D}_s$ depends on $\kappa$.

\subsection{Elliptical instability}
\label{sec:skyrmion:elliptical}

\begin{figure}
    \includegraphics[width=0.47\textwidth]{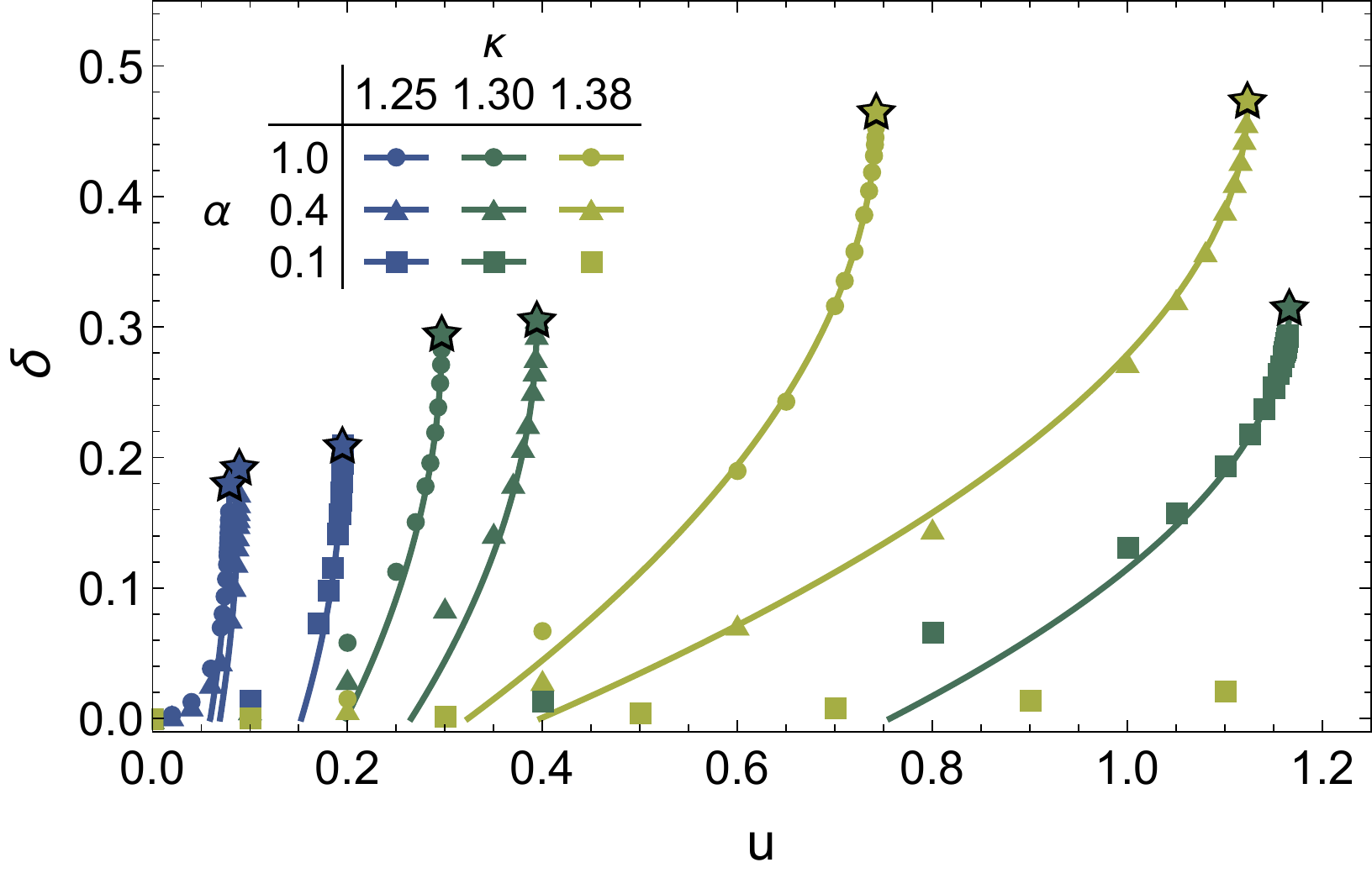}
    \caption{
    Distortion $\delta$ as function of the STT $u$ for $\kappa=1.25$ (blue), $\kappa=1.30$ (dark green), and $\kappa=1.38$ (light green), for $\alpha=1,0.4,0.1$.
    The numerical data (dots, triangles, squares) is supplemented by a $\sqrt{u_c-u}$ fit (solid line).
    The instability point is marked by a star.
    For $\kappa=1.38$ and $\alpha=0.1$ we do not find an instability of the skyrmion below the ferromagnetic instability.}
    \label{fig:skyrmion:elliptical:distortion}
\end{figure}

\begin{figure}    
    \includegraphics[width=\columnwidth]{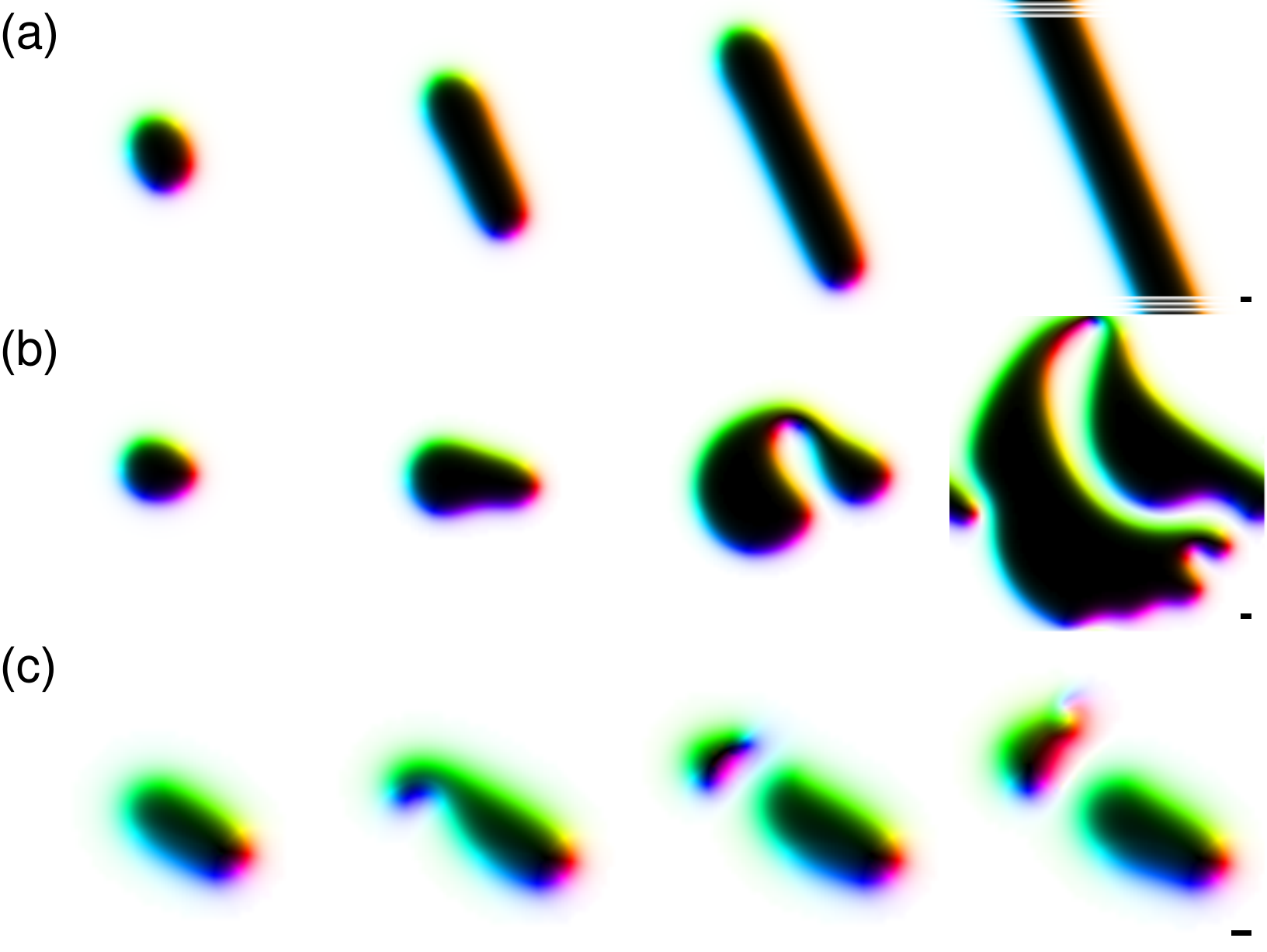}
    \caption{
        Time-evolution of the a variety of observed instabilities of a skyrmion due to STT: (a) elongation, (b) expansion, and (c) duplication.
        Parameters $(\kappa,\alpha,u)$ are (a) $(1.3,1.0,0.2977)$, (b) $(1.3,0.1,1.17)$, and (c) $(1.35,0.4,2.2)$.
        Instabilities (a) and (b) are obtained by increasing $u$ whereas (c) was initially above the unstable regime and $u$ was then decreased.
        For (a) we observed an expansion that was only limited by the size of the simulated area.
        In the last snapshot of (b), the skyrmion has filled the entire space over the periodic  boundary conditions.
        Both (b) and (c) result in unpredictable behavior at even longer timescales.
        The colorcode is chosen as in Fig.~\ref{fig:domainwalls:effect}.
        The current $\vect{u}=u\hat{\vect{x}}$ points to the right.
        The small black bar in the bottom right corner indicates the length scale $2A/D$ in each series of panels (panels (c) are magnified with respect to (a) and (b)).
    }
    \label{fig:skyrmion:elliptical:realspace}
\end{figure}

At larger effective drive $u$, beyond the linear regime and for smaller $\kappa\gtrsim\kappa_c$, the deformation parameter grows significantly, see Fig.~\ref{fig:skyrmion:elliptical:distortion}. For drives above a critical value, the steady solution is no longer possible and the skyrmion evolves dynamically at constant drive. This evolution corresponds to a growth in size that may lead to an elongation or the eventual break down into other structures, see Fig.~\ref{fig:skyrmion:elliptical:realspace}.

We associate the elongation behavior to an elliptical instability. This is due to the fact that the shape of the skyrmion resembles an ellipse with focal points distancing over time. The major bulk of the structure, however, can be compared to two parallel domain walls.
For such elongated magnetic structures, the distortion parameter reaches its limiting value of $\delta = 1$.
Besides the elongation instability, Fig.~\ref{fig:skyrmion:elliptical:realspace}(a), we also observe from numerical simulations other sorts of distortions associated to bending the skyrmion, Fig.~\ref{fig:skyrmion:elliptical:realspace}(b,c). 
While the elongated skyrmion tends to preserve mirror symmetry, this is lost for the other instabilities, which can be mainly associated to the field-like component of the effective field, see Eq.~\eqref{eq:EffectiveFieldGeneral}.

An analytical understanding of the elliptical instability can be obtained from Eq.~\eqref{eq:skyrmionradialprofile2}. The shape of the skyrmion is best described by its curvature $\Omega(l)$.
In the regime $\kappa \rightarrow \kappa_c$, and assuming that the profile along the skyrmion border is rather constant, i.e.\ $\partial_{l}\theta \approx 0$, (as consistent with micromagnetic simulations), the shape of the skyrmion for high drives resembles the ellipse
\begin{equation}
r(\psi) \approx \frac{R}{1 - e(v_{\mathrm{FL}},v_{\mathrm{DL}},R) R \cos\left(\theta_{\mathrm{DL}} - \psi\right)},
\end{equation}
hence the name ``elliptical'' instability.
 
An important remark is that the growth in shape of the skyrmion is associated with squeezing its radial profile, see Eq.~\eqref{eq:skyrmionradialprofile1}. In total, the skyrmion grows along $\vect{v}_{\textrm{DL}}$ and shrinks along the perpendicular direction.
Furthermore, we notice the existence of a critical behavior depending on the function $e(v_{\mathrm{FL}},v_{\mathrm{DL}},R)$, when $e(v_{\mathrm{FL}},v_{\mathrm{DL}},R) R = 1$. 
Numerical calculations show that this critical behavior happens for 
\begin{equation}
\label{eq:instabilitycondition}
e(v_{\mathrm{FL}},v_{\mathrm{DL}},R) R \approx \frac{v_{\mathrm{FL}}R^2}{2} \approx 1.
\end{equation} 
Above this limit, a solution for $\partial_{l}\theta \approx 0$ is no longer allowed such that the skyrmion becomes unstable or smoothly deforms into the ``shooting star" skyrmion, see Figs.~\ref{fig:skyrmion:phasediagrams:constK} and \ref{fig:skyrmion:phasediagrams:constAlpha}.

\subsection{Current-stabilized ``shooting star'' skyrmions}
\label{sec:skyrmion:shootingstars}
\begin{figure}[tb]
    \includegraphics[width=0.47\textwidth]{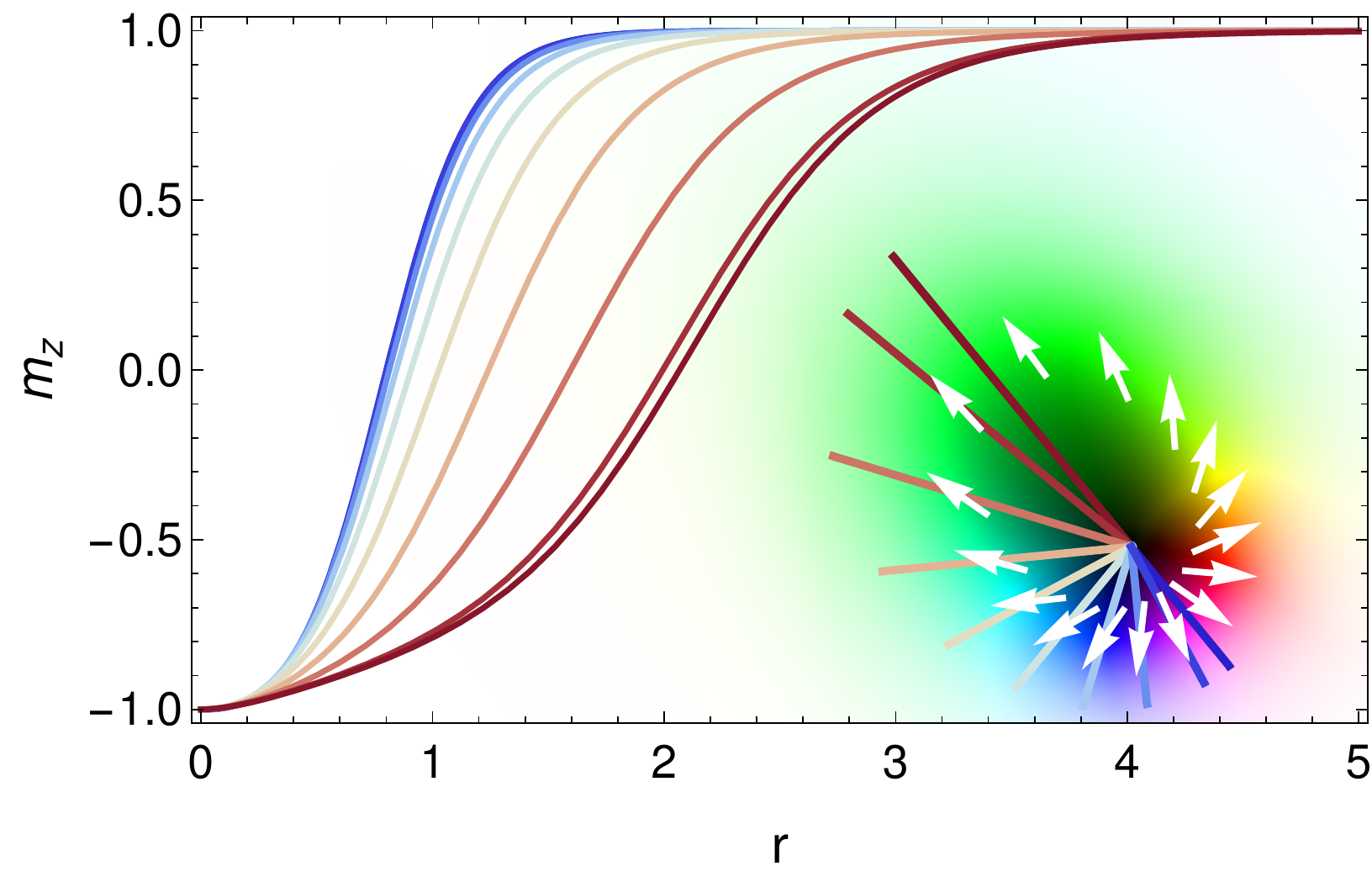}
    \caption{
        Profile of the ``shooting star'' skyrmion for $\kappa=1.4$ and $u=2.5$.
        The realspace magnetization for $\vect{u}=u\hat{x}$ is shown with the same colorcode as in Fig.~\ref{fig:domainwalls:effect}.
        Colored lines indicate the axes for which the $m_z$-component is plotted with the corresponding color.
        }
    \label{fig:skyrmion:shootingstars:profile}
\end{figure}

At high currents and larger damping parameters, a steady-state motion can be found where the shape of the magnetic skyrmion resembles a shooting star. 
This happens either for larger $\kappa$ or drives past the elongation instability, i.e., for parameter sets above the white regions in Figs.~\ref{fig:skyrmion:phasediagrams:constK} and~\ref{fig:skyrmion:phasediagrams:constAlpha} (b,c). 
An example of such a shooting-star like skyrmion is displayed in the inset of Fig.~\ref{fig:skyrmion:shootingstars:profile}, alongside with its radial profile, which changes significantly as indicated by the different colors. 
The stability of the shooting star skyrmion for such high drives can be explained by the fact that the energy associated to the skyrmion profile deformation is much higher than the energy associated with the eigenmodes that lead to its boundary contour change. \cite{Rodrigues2018,Kravchuk2018}
We observe that these shooting star skyrmions are very rigid structures, even at ultra-high currents. 
Upon increasing the drive in this phase, we find that they become even more compact.
 
The numerics indicate that the shooting star solutions are rather mirror symmetric with respect to the axis $\vect{v}_{\mathrm{DL}}$.
As stated above and proven in App.~\ref{app:symmetry}, for such mirror symmetric solutions 
i) $\vect{v}_{\mathrm{DL}} \perp \vect{v}_{\mathrm{FL}}$, 
ii) $\vect{v}_{\mathrm{DL}}$ and $\vect{v}_{\mathrm{FL}}$ are the eigenvectors of $\mathcal{D}$, and 
iii) the axis of distortion is along $\vect{v}_{\mathrm{DL}}$.

\section{Current-induced deformation of skyrmions stabilized by magnetic field}
\label{sec:bfield} 

\begin{figure}
    \includegraphics[width=0.336\textwidth]{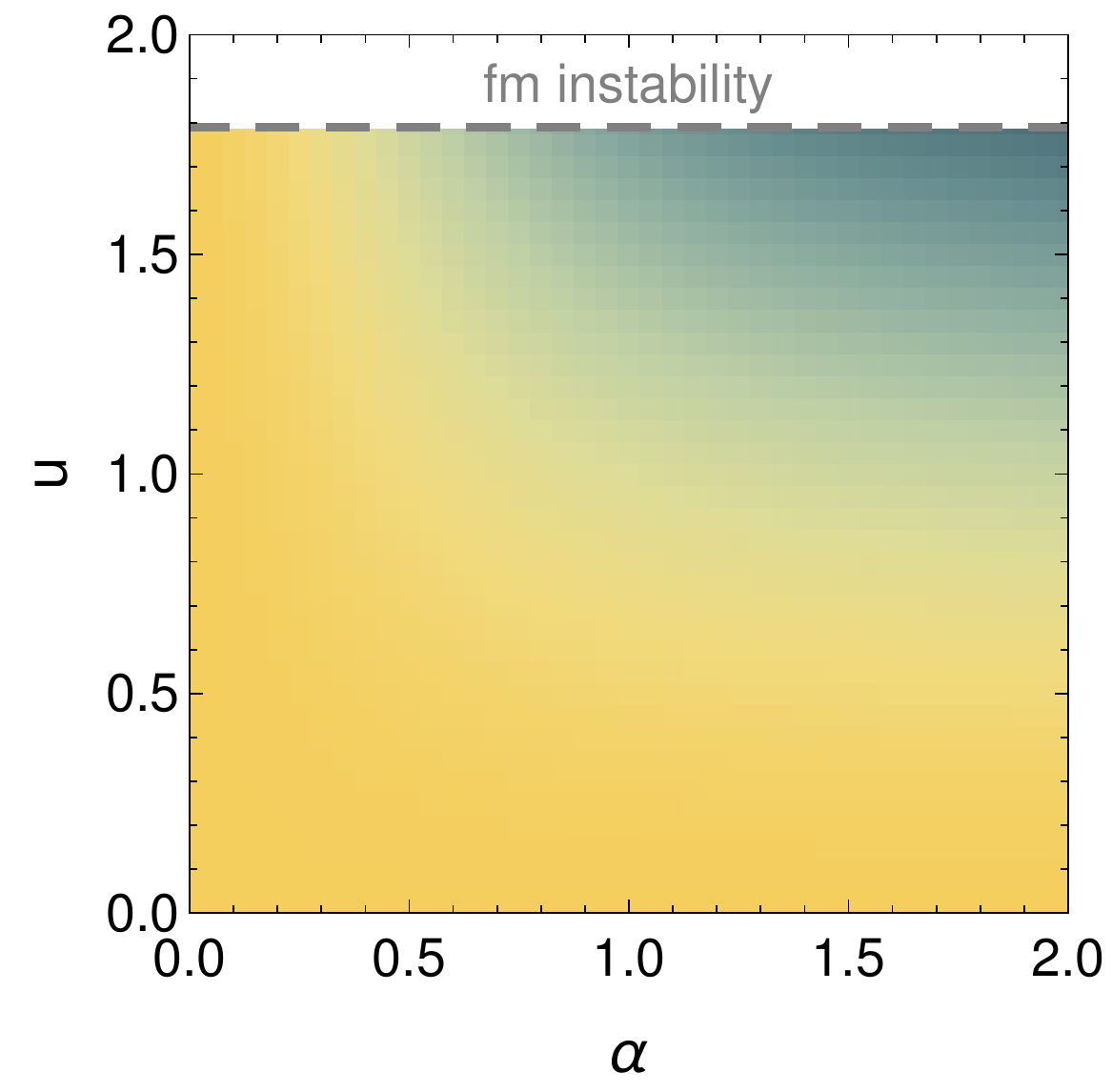}
    \hspace{0mm}
    \includegraphics[width=0.065\textwidth]{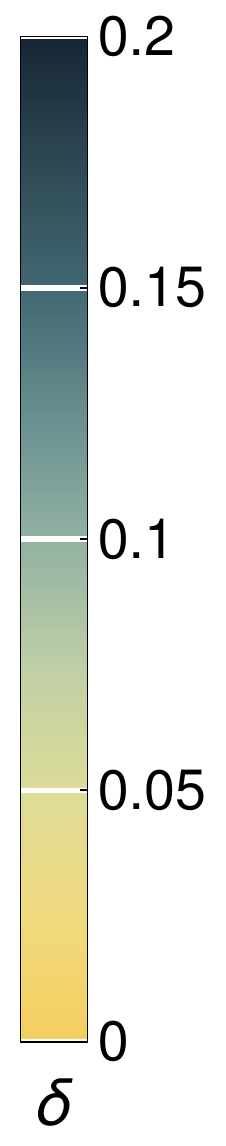}
    \caption{
        Distortion $\delta$ of skyrmions stabilized by a magnetic field $h = 0.8$ (in rescaled units) without uniaxial anisotropy as function of Gilbert damping $\alpha$ and drive $u$.
        For better visibility, the color code only ranges from $\delta=0$ (yellow) to $\delta=0.2$ (dark green), see colorbar.
        The solid gray line shows the ferromagnetic instability at $u=2 \sqrt{h}$ above which the ferromagnetic background becomes unstable, c.f. Sec.~\ref{sec:model:ferromagnet}.  }
    \label{fig:appendix:bfield}
\end{figure}

So far, we only considered skyrmions which are stabilized in the absence of external magnetic fields and, instead, by a uniaxial anisotropy.
However, stable skyrmion solutions can also be found for additional external magnetic fields, even for vanishing or in-plane anisotropy.~\cite{Heinze2011,Leonov2015b,Leonov2016a}
Yet the thermodynamic phase transitions, as well as the energetics of excitations of a skyrmion in the two regimes -- field vs. anisotropy-stabilized -- are very different:
As discussed above, anisotropy-stabilized skyrmions become very soft close to the phase transition to the spiral phase and can easily deform under an applied currents.
For magnetic field-stabilized skyrmions in the absence of a uniaxial anistropy, however, we find that the skyrmion is rather rigid in the proximity of the phase transition. 
It shows only little deformation due to STTs over a wide range of parameters $\alpha$ and $u$, as shown in Fig.~\ref{fig:appendix:bfield}.
This can be explained by the following argument: 
The ferromagnet becomes unstable against the formation of a skyrmion lattice at the critical field $h_{c_1}$ which in rescaled units is given by $h_{c_1}. \approx 0.8$\cite{Bogdanov1989,Romming2015,Gilbert2015, Garst2017} 
However, the instability of the skyrmion in a polarized background occurs via the softening of the elliptical excitation mode at much lower fields $h_{c_2} \approx 0.56$.\cite{Ezawa2011,Schutte2014,Lin2014} 
Therefore, excitations are massively gapped around the phase transition \cite{Lin2014,Schutte2014,Garst2017} and the skyrmion cannot deform much.

\section{Discussion and conclusions}
\label{sec:conclusions}

In this paper we have systematically investigated the properties of isolated skyrmions in steady-state motion, including their deformations and instabilities, in chiral ferromagnets due to STTs. 
In general, we have found that skyrmions deform away from their typical circular shape for applied drives and we have quantified their distortion of the skyrmion by the single scalar parameter $\delta$, see Eq.~\eqref{eq:skyrmion:distortion}.
As depicted in Figs.~\ref{fig:skyrmion:phasediagrams:constK} and \ref{fig:skyrmion:phasediagrams:constAlpha},
depending on the size of a skyrmion and the damping parameter, there are various scenarios which happen as a function of driving strength: 
i) For a smaller skyrmion at smaller damping, it remains rather circular; 
ii) for a larger skyrmion at smaller damping, it first develops an elliptical instability which then goes over into the ferromagnetic instability;
iii) for a larger skyrmion at larger damping, it first develops an elliptical instability which then contracts back to a shooting-star shaped form;
or iv) for a smaller skyrmion at larger damping, it transforms continuously into the shooting-star skyrmion solution.
In particular, in this work we have predicted the shooting star skyrmion as a new, less symmetric state which is stable at ultra-high currents. 
This is due to the fact that its radial profile varies significantly (unlike for a usual skyrmion).

Even though we have derived our results explicitly for interfacial DMI, we would like to point out that our results are independent of the flavor of DMI, as long as its absolute value is isotropic.
By replacing $\hat{z}\!\times\!\nabla \to \mathcal{R}_z^\phi\nabla$ where $\mathcal{R}_z^\phi$ is a rotation by $\phi$ around the $\hat{z}$-axis, any other plane of rotation can be stabilized, including Bloch-type skyrmions.
Note that also anti-skyrmions can be considered, as $\hat{z}\!\times\!\nabla \to \mathcal{R}_z^\phi\sigma_x\nabla$ (where $\sigma_x$ is a Pauli matrix).

To summarize, our results show that uniaxial anisotropy-stabilized skyrmions are in fact rather soft compared to magnetic field stabilized skyrmions.
Hence, for any technological devices aiming to exploit skyrmions in chiral magnets with an easy-axis anisotropy, in the absence of invasive external magnetic fields, it is essential to know the range of parameters where the skyrmion is unstable.
Otherwise, the spin-torques can destroy of the skyrmion or, at least, modify its shape and hence skyrmion Hall effect which then leads to a less controlled motion.


\section{Acknowledgments}
J.M. acknowledges the fruitful discussions with A. Rosch, K. Litzius, N. Nagaosa, and C. Melcher.
J.M. was supported in Germany by the German Research Foundation (DFG) CRC 1238 project C04 and in Japan by JSPS (project No. 19F19815) and the Alexander von Humboldt foundation.
The group at Mainz acknowledges funding from the German Research Foundation (DFG) under the Project No.~EV 196/2-1, EV196/5-1, SI1720/4-1, TRR173 - 268565370 (project B11), the Graduate School of Excellence Materials Science in Mainz (MAINZ, GSC 266), and from the Emergent AI Center funded by the Carl-Zeiss-Stiftung.  
We furthermore thank the Regional Computing Center of the University of Cologne (RRZK) for providing computing time on the DFG-funded High Performance Computing (HPC) system CHEOPS as well as support.


\appendix

\section{Simulation details}
\label{sec:appendix:simulationdetails}

\subsection{Numerical implementation of the continuum model}
\label{sec:appendix:simulationdetails:numerical}

Minimizing artifacts due to numerical discretization of the continuum theory, Eq.~\eqref{eq:model:energy} and Eq.~\eqref{eq:model:LLG}, requires a numerical grid with a small lattice constant $a\ll\xi$, where $\xi$ is the typical length scale of variations of the magnetization.
The large skyrmions at low $\kappa$ are particularly sensitive to the anisotropies that arise from the numerical discretization of the grid. This erroneously leads to non-circular skyrmions already without applied currents and wrong results for the axis of distortion.
With the standard micromagnetic solvers and their $\mathcal{O}(a^2)$ finite difference schemes, a very fine discretization is therefore required to reduce these effects which leads to very long runtimes.

Therefore, in this work, we use higher order stencils with a numerical error of the order $\mathcal{O}(a^8)$ for the discretization of derivatives, following the continuum calculations in Ref.~\onlinecite{Heil2019}.
Because of the high order scaling of the numerical error, a square lattice with lattice constant $a=\frac{1}{3}\frac{2A}{D}$ or $a=\frac{1}{6}\frac{2A}{D}$ was found to be sufficiently precise, and thus, was used throughout this work.
The usual lattice sizes are then only $60\times60$, $100\times100$, or $200\times200$, depending on the size of the skyrmion.

\subsection{Simulations of moving skyrmions in Thiele's comoving frame of reference}
\label{sec:appendix:simulationdetails:comovingframe}

As we apply a strong current to our system, the skyrmion moves very rapidly over the numerical lattice.
To avoid numerical precision errors, we simulate the system in a frame of reference which moves with the magnetic texture.
Instead of numerically tracking the position of the skyrmion, which is hard to do at high precision, we estimate the velocity $\vect{v}_\mathrm{sky}$ of the skyrmion from the Thiele equation, Eq.~\eqref{eq:skyrmion:thiele}.
To account for the changes in the magnetic texture during the simulation, the dissipation matrix elements become time-dependent, $\mathcal{D}\to\mathcal{D}(t)$, and have to be evaluated at each time-step of the simulation.
We can write the time-dependent magnetization in the comoving frame of the skyrmion as $\vect{n} (\vect {r}, t) = \vect{m} (\vect {r} - \vect{R}_\mathrm{sky}(t), t)$
where $\vect{R}_\mathrm{sky}(t)$ is the position of the skyrmion with $\dot{\vect{R}}_\mathrm{sky}(t) = \vect{v}_\mathrm{sky}(t)$, see Eq.~\eqref{eq:skyrmion:thiele}.
The resulting LLG equation is
\begin{equation}
\begin{split}
\dot {\vect{n}} =\,
    &\frac{1}{1+\alpha^2} \Big(- \gamma\, \vect{n} \!\times\! \vect{B}_{\mathrm{eff}} - (1\!+\!\alpha\beta)\left(\vect{v}_e \!\cdot\! \nabla\right) \vect{n} \\
    &- \alpha\gamma \, \vect{n} \!\times\! (\vect{n} \!\times\! \vect{B}_{\mathrm{eff}}) - (\alpha\!-\!\beta)\, \vect{n}\!\times\!( \vect{v}_e \!\cdot\! \nabla ) \vect{n} \Big) \\
    &+ (\vect{v}_\mathrm{sky}(t)\!\cdot\!\nabla)\,\vect{n},
 \end{split}
\label{eq:appendix:simulationdetails:comovingframe:comovingLLG}
\end{equation}
where $\vect{v}_\mathrm{sky}(t)$ is calculated at each time-step. Note that the equation has been written in a form with the time-derivative appearing only on the left hand side, making it suitable for direct numerical integration.

Moreover, as we study the steady motion of skyrmions, we have to apply the current quasi-adiabatically, avoiding sharp accelerations of the skyrmion.
In the simulations this is achieved via smoothly increasing the current density to its final strength $ v_\mathrm{f}$ with a time-dependence $v(t) = v_\mathrm{f} \sin^2(t/t_0)$ where $t_0$ is chosen sufficiently large such that excitations of the skyrmion are avoided. A typical value of $t_0$ that we have used is $t_0 \approx 6\cdot10^4$ in dimensionless units.
The time-integration of the co-moving LLG equation, Eq.~\eqref{eq:appendix:simulationdetails:comovingframe:comovingLLG}, is then performed until a steady state is reached.

\section{Proof that the effective velocites $\vect{v}_{\mathrm{FL}}$ and $\vect{v}_{\mathrm{DL}}$ are always nonzero for a skyrmion subject to finite drive $\vect u$}
\label{app:v1-v2-analysis}

Let us assume that $\vect{v}_{\mathrm{DL}}=\vect{0}$, i.e.\ the damping-like part of Eq.~\eqref{eq:skyrmion:movingLLG} equals zero.
In this case from Eq.~\eqref{eq:skyrmion:thiele} we obtain
\begin{align}
-\vect{u}
&=
	\frac{\alpha}{\mathcal{G}^{2}+\alpha^{2}\det(\mathcal{D})}
\left(
	\vect{\mathcal{G}}
	\times
	(\mathcal{D}\vect{u})
	-\alpha\det(\mathcal{D})\vect{u}
\right)
\end{align}
which reduces to an eigenvalue equation
\begin{align}
G\mathcal{D}\vect{u}
&=
	-(\mathcal{G}^{2}/\alpha)\vect{u},  \quad  \text{where} \quad 
	G=
\begin{pmatrix}
0 & -\mathcal{G} & 0 \\
\mathcal{G} & 0 & 0 \\
0 & 0 & 1
\end{pmatrix}.
\end{align}
For a non-trivial solution to this equation, $\vect{u}$ has to be an eigenvector of $G\mathcal{D}$ with eigenvalue $-(\mathcal{G}^{2}/\alpha)$.
However, explicitly calculating the eigenvalues of $G\mathcal{D}$ we obtain $ \pm |\mathcal{G}|\sqrt{-\det\mathcal{D}}$, which for an arbitrary skyrmion leads to a contradiction.

Let us now assume that $\vect{v}_{\mathrm{FL}}=\vect{0}$, leaving just the purely damping-like spin-torque in Eq.~\eqref{eq:skyrmion:movingLLG}.
By a similar analysis, from Eq.~\eqref{eq:skyrmion:thiele} we obtain the eigenvalue equation for $\vect u$%
\begin{align}
G\mathcal{D}\vect{u} = \alpha\det(\mathcal{D}) \vect{u},
\end{align}
which for a general skyrmion can also not be fulfilled.

To conclude, for any finite $u > 0$, and thus \ $\alpha \neq \beta$, both $\vect{v}_\mathrm{FL}$ and $\vect{v}_\mathrm{DL}$ are always simultaneously non-zero, as stated in the main text.

\section{Detailed results for the distortion}
\label{app:phasediagramdetails}

\begin{figure}[tb]

    \includegraphics[width=0.47\textwidth]{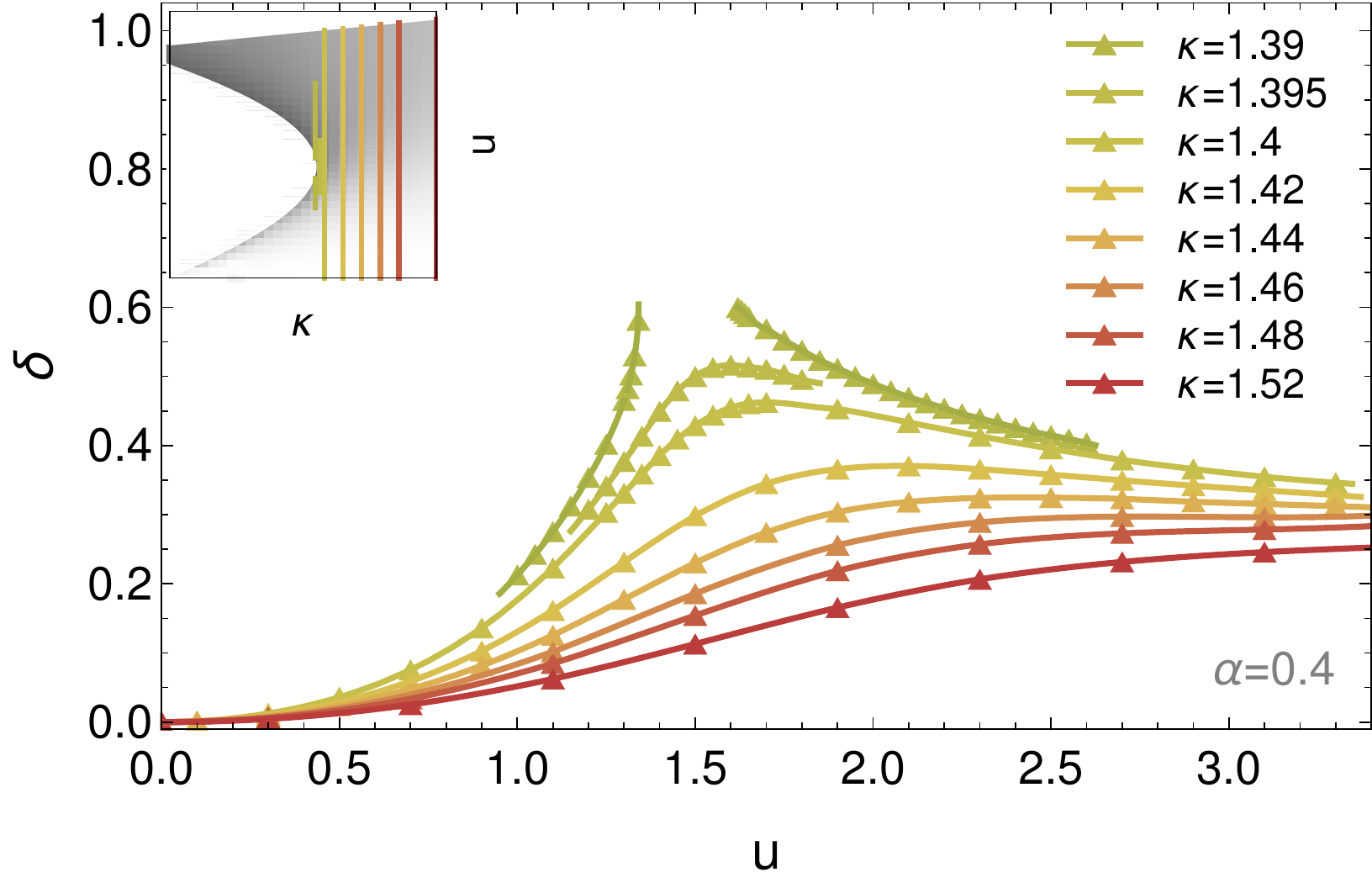}
    
    \includegraphics[width=0.47\textwidth]{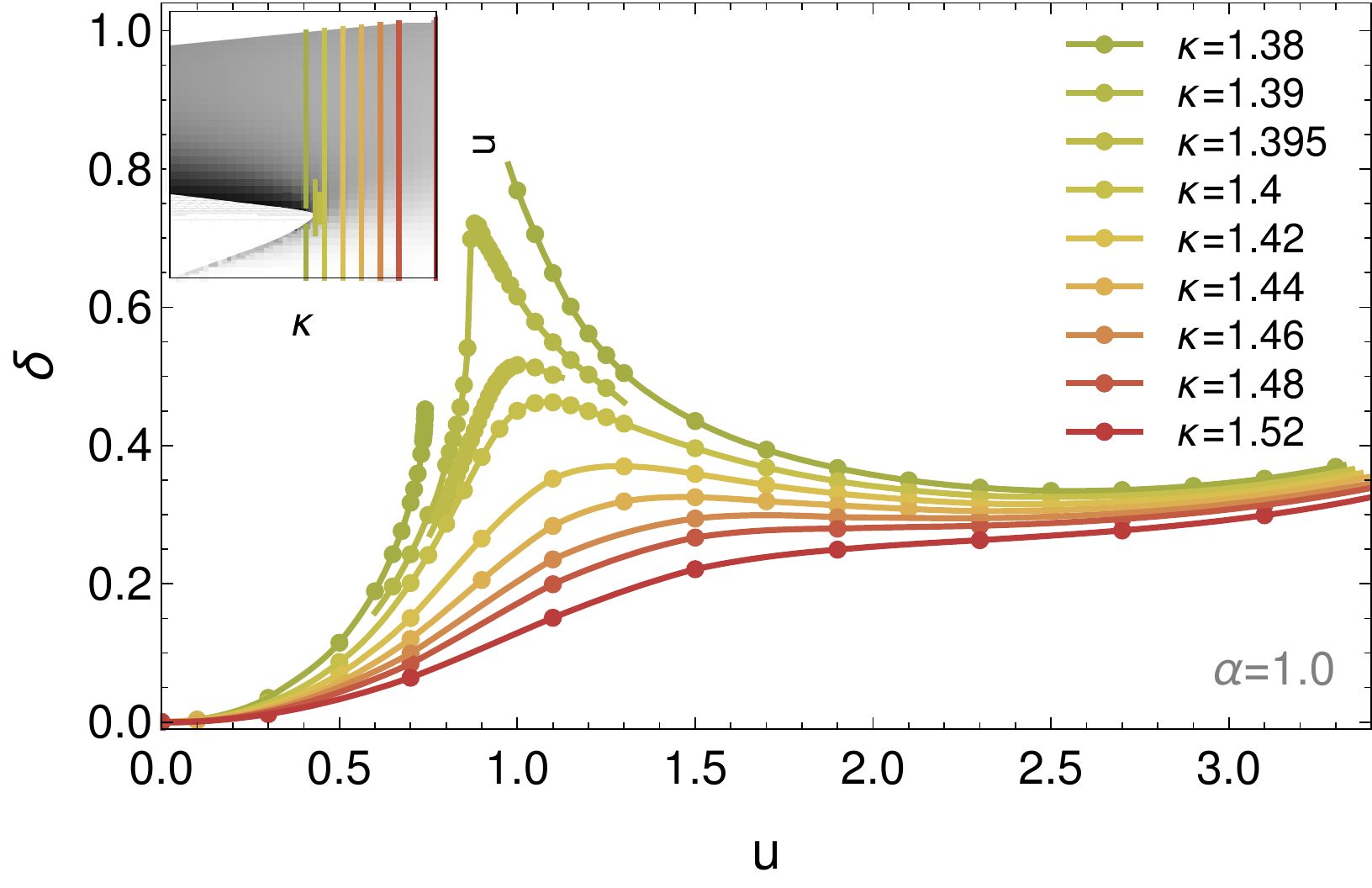}
   
   \caption{
        Distortion $\delta$ as function of applied drive $u$ for damping $\alpha=0.4$ and $\alpha=1$.
        Dots/triangles indicate numerically obtained results and lines are interpolations.
        The color denotes the anisotropy $\kappa$.
        The inset is a reproduction of the full $\kappa$-$u$ phase diagrams in gray scale, c.f. Fig.~\ref{fig:skyrmion:phasediagrams:constAlpha}, where the horizontal lines correspond to the data shown.
   }
   \label{fig:detaileddistortion}
\end{figure}

In Fig.~\ref{fig:detaileddistortion} we plot the detailed numerical results for the deformation parameter shown in Figs.~\ref{fig:skyrmion:phasediagrams:constAlpha}(b,c) in the main text. 
These data explicitly display the different behavior of the distortion of the different phases -- slightly deformed skyrmions, instability phase and the shooting star.
For the lowest $\kappa$ shown here, the skyrmion exhibits an instability for an intermediate range of $u$.
At higher $\kappa$, the unstable regime closes but a peak is still visible which marks the transition to a new current-stabilized skyrmion state which we refer to as the ``shooting star'' skyrmion.
For even higher anisotropy this peak vanishes and the transition to the shooting star becomes a smooth crossover.

\section{Mirror symmetric current-driven skyrmions}
\label{app:symmetry}

First we show that, if the skyrmion is mirror symmetric around a certain axis, then, in the coordinate system where this axis is along one of the  basis vectors, the dissipative matrix is diagonal. 
To do that, we consider without loss of generality $\hat{\vect{x}}$ as the axis of the mirror symmetry, such that $m_{x} (x,y) = m_{x}(x,-y)$ and $m_{y} (x,y) = -m_{y}(x,-y)$. 
It follows that $\mathcal{D}_{xy} = \int d^2 x\, \partial_{x}\vect{m}(x,y)\cdot \partial_{y}\vect{m} (x,y) = 0$, since $\partial_{y}\vect{m} (x,-y) = - \partial_{y}\vect{m} (x,y)$ and $\partial_{x}\vect{m} (x,-y) = \partial_{x}\vect{m} (x,y)$. 

Second, we show that if the current-driven skyrmion is mirror symmetric with respect to $\vect{v}_{\mathrm{DL}}$ then, and only then, $\vect{v}_{\mathrm{DL}} \perp \vect{v}_{\mathrm{FL}}$. To proof this, we consider Eq.~\eqref{eq:skyrmion:movingLLG} in a coordinate frame such that $\hat{\vect{x}} \parallel \vect{v}_{\mathrm{DL}}$. In this coordinate frame, we can express  $\vect{v}_{\mathrm{FL}} $ as 
$\vect{v}_{\mathrm{FL}} = v_{\mathrm{FL}\,x} \hat{\vect{x}} + v_{\mathrm{FL}\,y} \hat{\vect{y}}$ where, $v_{\mathrm{FL}\,x} = \vect{v}_{\mathrm{FL}}\cdot\vect{v}_{\mathrm{DL}}$, the component of $\vect{v}_{\mathrm{FL}}$ along $\vect{v}_{\mathrm{DL}}$ and $v_{\mathrm{FL}\,y}$ the component of $\vect{v}_{\mathrm{FL}}$  perpendicular to $\vect{v}_{\mathrm{DL}}$. In this case, Eq.~\eqref{eq:skyrmion:movingLLG} reduces to
\begin{equation}
 \vect{0}
=  \vect{m} \times \left[ \vect{B}_\mathrm{eff} +
\alpha v_\mathrm{DL}\partial_{x} \vect{m}\right] - \left(v_{\mathrm{FL}\,x} \partial_{x} + v_{\mathrm{FL}\,y} \partial_{y}\right) \vect{m}.
\end{equation}
If we project the equations above along $\vect{m}\times\partial_{x}\vect{m}$ and $\vect{m}\times\partial_{y}\vect{m}$ and integrate over space we obtain the following Thiele equations, respectively,
\begin{subequations}
\begin{align}
\alpha \, v_\mathrm{DL} \mathcal{D}_{xx} - 4\pi \mathcal{Q} v_\mathrm{FL\,y} = 0,\\
\alpha \, v_\mathrm{DL} \mathcal{D}_{xy} + 4\pi \mathcal{Q} v_\mathrm{FL\,x} = 0.
\end{align}
\end{subequations}
Here we used that the energy, Eq.~\eqref{eq:model:energy}, is translationally invariant, i.e.\ $\delta_{x} E[\vect{m}]  = \delta_{y} E[\vect{m}] = 0$. The only solution for this system of equations such that both effective drives $\vect{v}_\mathrm{DL}$ and $\vect{v}_{\mathrm{FL}}$ are non-zero and the dissipative matrix $\mathcal{D}$ is diagonal, is given by $v_\mathrm{FL\,x} = 0$ and $|\vect{v}_{\mathrm{FL}}| = \alpha |\vect{v}_{\mathrm{DL}}|  \mathcal{D}_{xx}/ 4\pi \mathcal{Q}$.
Moreover, then $\vect{v}_{\mathrm{DL}}$ and $\vect{v}_{\mathrm{FL}}$ are the eigenvectors of the dissipation matrix $\mathcal{D}$ and $\vect{v}_{\mathrm{DL}}$ is along the axis of distortion. 
Notice that in a different basis, the mirror symmetric dissipation matrix components become $\mathcal{D}_{x'x'} = \mathcal{D}_{xx}\cos^2\theta' + \mathcal{D}_{yy}\sin^2\theta'$, $\mathcal{D}_{y'y'} = \mathcal{D}_{xx}\sin^2\theta' + \mathcal{D}_{yy}\cos^2\theta'$, and $\mathcal{D}_{x'y'} = (\mathcal{D}_{yy} - \mathcal{D}_{xx})\sin\theta'\cos\theta'$, where $\theta'$ is the angle between $\vect{v}_{\mathrm{DL}}$ and $\hat{\vect{x}}'$.

\section{Derivation for skyrmion deformation in the limit of low drive}
\label{app:lowdrive}

A quantitative analysis of the deformation $\delta$ can be obtained by a linearization of Eqs.~\eqref{eq:skyrmionradialprofile} on the drive parameter $u$. We take into account that up to linear order in the drive $\vect{v}_{\mathrm{DL}} \perp \vect{v}_{\mathrm{FL}}$, i.e.\ $\theta_{\mathrm{FL}} = \theta_{\mathrm{DL}} - \pi/2$. Furthermore, we consider a perturbative ansatz around the circular skyrmion solution with radius $R$ characterized by $\phi_0=0$ and $\theta_0(n)$. The latter solves 
the Eqs.~\eqref{eq:skyrmionradialprofile} for a circular skyrmion which reduce to
\begin{equation}\label{eq:RadialProfCircSkyrmion}
\partial_{n}^2\theta_{0} + \frac{2}{R}\sin^2\theta_{0}
- \frac{\sin2\theta_{0}}{2}\left(2\kappa + \frac{1}{R^2}\right) = 0.
\end{equation}
The perturbative ansatz is given by
\begin{subequations}
\begin{align}
\theta(n,l) &= \theta_{0}(n) + \tilde{\theta}(n,l)\\ 
\phi(n,l) &= \tilde{\phi}(l),
\end{align}
\end{subequations}
 where the functions with the tilde are small perturbations.
With this ansatz follows, $r(l) = R + \tilde{r}(l)$, $\Omega(l) = (1/R) + \tilde{\Omega}(l)$, with $\tilde{\Omega}(l) = \partial_{l}\tilde{\phi} -(\partial_{l}^2\tilde{r} + \tilde{r})/R^2$, $\Theta = \psi - \partial_{\psi}\tilde{r}/R$, and $\partial_{l} = (1/R - \tilde{r}/R^2)\partial_{\psi}$ where $\psi$ is the polar angle. This ansatz simplifies Eqs.~\eqref{eq:skyrmionradialprofile} to the following system of equations up to first order in perturbation,
\begin{subequations}\label{eq:skyrmionradialprofilelinear}
\begin{align}\label{eq:skyrmionradialprofilelinear1}
0=\ &\partial_{n}^2\tilde{\theta} + \partial_{l}^2\tilde{\theta} + 2\tilde{\Omega}\sin^2\theta_{0} + 2\frac{\sin2\theta_{0}}{R}\tilde{\theta}\\
&- \frac{\sin2\theta_{0}}{R}\tilde{\Omega} - \tilde{\theta}\cos2\theta_{0}\left(2\kappa + \frac{1}{R^2}\right) \notag\\
&+\cos(\theta_{\mathrm{DL}} - \psi)\left(\alpha v_{\mathrm{DL}} \partial_{n}\theta_{0} + \frac{v_{\mathrm{FL}}\sin\theta_{0}}{R}\right),\notag\\[0.2mm]
\label{eq:skyrmionradialprofilelinear2}
0=\ & \frac{2\partial_{l}\tilde{\theta}\cos\theta_{0}}{R} +  \sin\theta_{0}\left(2\partial_{n}\theta_{0}\tilde{\phi} + \partial_{l}\tilde{\Omega}\right)\\
&+ \sin(\theta_{\mathrm{DL}} - \psi)\left( v_{\mathrm{FL}}\partial_{n}\theta_{0} + \frac{\alpha v_{\mathrm{DL}}\sin\theta_{0}}{R}\right).\notag
 \end{align}
 \end{subequations}
  
From the radial profile \eqref{eq:RadialProfCircSkyrmion}, in the limit of big radius, we can consider that $\partial_{n}\theta_{0} \sim f(\kappa)\sin\theta_{0}/R$ such that solutions of Eqs.~\eqref{eq:skyrmionradialprofilelinear} can be approximated by functions of $(\alpha v_{\mathrm{DL}} - v_{\mathrm{FL}})\cos(\theta_\mathrm{DL} - \psi)$ and  $(\alpha v_{\mathrm{DL}} - v_{\mathrm{FL}})\sin(\theta_\mathrm{DL} - \psi)$. By expanding the components of the dissipation matrix, Eq.~\eqref{eq:DMatrixDW}, into the lowest order of the perturbation, considering the basis with $\hat{\vect{x}} \parallel \vect{v}_\mathrm{DL}$, we obtain: i) $\mathcal{D}_{xy} = 0$, such that the skyrmion is mirror symmetric with respect to $\vect{v}_{\textrm{DL}}$; and ii) the first non-zero contribution to the distortion is given by
\begin{equation}
\delta \propto \left((\alpha v_{\mathrm{DL}} - v_{\mathrm{FL}})^2 - c(\kappa/\kappa_{c}) \alpha v_{\mathrm{DL}}v_{\mathrm{FL}}\right)/R^2,
\end{equation} 
where $c(\kappa/\kappa_{c})$ is a function of $\kappa$ and depend on the exact solutions of Eqs.~\eqref{eq:skyrmionradialprofilelinear}, as claimed in the main text.

%

\end{document}